 \documentclass[preprint2]{aastex}

\slugcomment{Not to appear in Nonlearned J., 45.}

\shorttitle{X-RAY GROUPS OF GALAXIES IN AEGIS FIELD}
\shortauthors{Erfanianfar et al.}

\begin{document}


\title{X-RAY GROUPS OF GALAXIES IN THE AEGIS DEEP AND WIDE FIELDs}

\author{G. Erfanianfar\altaffilmark{1},
A. Finoguenov\altaffilmark{2,3}, 
M. Tanaka\altaffilmark{4},
M. Lerchster\altaffilmark{1},
K. Nandra\altaffilmark{1},
E. Laird\altaffilmark{5}, 
J. L. Connelly\altaffilmark{1}, 
R. Bielby\altaffilmark{6,7}, 
M. Mirkazemi\altaffilmark{1}, 
S. M. Faber\altaffilmark{8}, 
D. Kocevski\altaffilmark{8}, 
M. Cooper\altaffilmark{9,\dagger}, 
J. A. Newman\altaffilmark{10}, 
T. Jeltema\altaffilmark{8,11}, 
A. L. Coil\altaffilmark{12}, 
F. Brimioulle\altaffilmark{13}, 
M. Davis\altaffilmark{14}, 
H. J. McCracken\altaffilmark{7}, 
C. Willmer\altaffilmark{15}, 
B. Gerke\altaffilmark{16}, 
N. Cappelluti\altaffilmark{17}, 
S. Gwyn\altaffilmark{18} }

\email{erfanian@mpe.mpg.de}

\altaffiltext{1}{Max Planck-Institute for extraterrestrial Physics, PO Box 1312, Giessenbachstr. 1., 85741 Garching, Germany}
\altaffiltext{2}{Department of Physics, University of Helsinki, Gustaf H\"allstr\"omin katu 2a, FI-00014 Helsinki, Finland}
\altaffiltext{3}{University of Maryland, Baltimore County, 1000 Hilltop Circle, Baltimore, MD 21250, USA}
\altaffiltext{4}{Institute for the Physics and Mathematics of the Universe, The University of Tokyo, 5-1-5 Kashiwanoha, Kashiwa-shi, Chiba 277-8583, Japan}
\altaffiltext{5}{Astrophysics Group, Imperial College London, Blackett Laboratory, Prince Consort Road, London SW7 2AZ, UK}
\altaffiltext{6}{Department of Physics, Durham University, South Road, Durham, DH1 3LE, UK}
\altaffiltext{7}{Institut d'Astrophysique de Paris, UMR 7095 CNRS, Université Pierre et Marie Curie, 98 bis boulevard Arago, 75014 Paris, France}
\altaffiltext{8}{UCO/Lick Observatories, Department of Astronomy and Astrophysics, University of California, Santa Cruz, CA 95064, USA}
\altaffiltext{9}{Center for Galaxy Evolution, Department of Physics and Astronomy, University of California, Irvine, 4129 Frederick Reines Hall, Irvine, CA 92697, USA} 
\altaffiltext{$\dagger$}{Hubble Fellow}
 \altaffiltext{10}{Department of Physics and Astronomy, University of Pittsburgh, 401-C Allen Hall, 3941 O'Hara Street, Pittsbrugh, PA 15260, USA}
\altaffiltext{11}{Department of Physics and Santa Cruz Institute for Particle Physics, University of California, 1156 High St., Santa Cruz, CA 95064, USA}
\altaffiltext{12}{Center for Astrophysics and Space Sciences, University of California, San Diego, 9500 Gilman Drive, MC 0424, San Diego, CA 92093, USA}
 \altaffiltext{13}{University Observatory Munich, Ludwigs-Maximilians University Munich, Scheinerstr. 1, 81679 Munich, Germany}
 \altaffiltext{14}{Department of Astronomy, University of California, Berkeley, Berkeley, CA 94720, USA}
 \altaffiltext{15}{Steward Observatory, University of Arizona, 933 N. Cherry Avenue, Tucson, AZ 85721, USA}
 \altaffiltext{16}{Lawrence Berkeley National Laboratory, 1 Cyclotron Rd. MS 90R4000, Berkeley, CA 94720}
 \altaffiltext{17}{INAF-Osservatorio Astronomico di Bologna, via Ranzani 1, 40127 Bologna, Italy}
\altaffiltext{18}{Canadian Astronomy Data Centre, Herzberg Institute of Astrophysics, 5071 West Saanich Road, Victoria, British Columbia, V9E 2E7, Canada}

\begin{abstract}
We present the results of a search for extended X-ray sources and their corresponding galaxy groups from 800-ks Chandra coverage
 of the All-wavelength Extended Groth Strip International Survey (AEGIS). This yields one of the largest
X-ray selected galaxy group catalogs from a blind survey to date. The red-sequence technique and spectroscopic redshifts allow
 us to identify 100$\%$  of reliable sources, leading to a catalog of 52 galaxy groups. The groups span the redshift range
 $z\sim0.066-1.544$ and virial mass range $M_{200}\sim1.34\times 10^{13}-1.33\times 10^{14}M_\odot$. For the 49
 extended sources which lie within DEEP2 and DEEP3 Galaxy Redshift Survey coverage, 
we identify spectroscopic counterparts and determine velocity dispersions. We select member galaxies by applying different cuts along the line of 
sight or in projected spatial coordinates. A constant cut along the line of sight can cause a large scatter in scaling relations in low-mass or high-mass
 systems depending on the size of cut. A velocity dispersion based virial radius can more overestimate velocity dispersion in 
comparison to X-ray based virial radius for low mass systems. There is no significant difference between these two radial cuts for more massive systems.
Independent of radial cut, overestimation of velocity dispersion can be created in case of existence of significant substructure and also compactness in X-ray 
emission which mostly occur in low mass systems. We also present a comparison between X-ray galaxy groups and optical galaxy groups 
detected using the Voronoi-Delaunay method (VDM) for DEEP2 data in this field.
\end{abstract}


\section{Introduction}
Groups of galaxies are important laboratories to study galaxy evolution and formation. They are in the stage between the field and the densest 
environment in the universe, massive clusters (\citealt{Zab98}) and as many as $50\%-70\%$ of all galaxies reside in galaxy groups 
\citep{Tur76,Gel83,Eke05}. 
\begin{figure*}[t]
\centering

 \includegraphics[width=12cm]{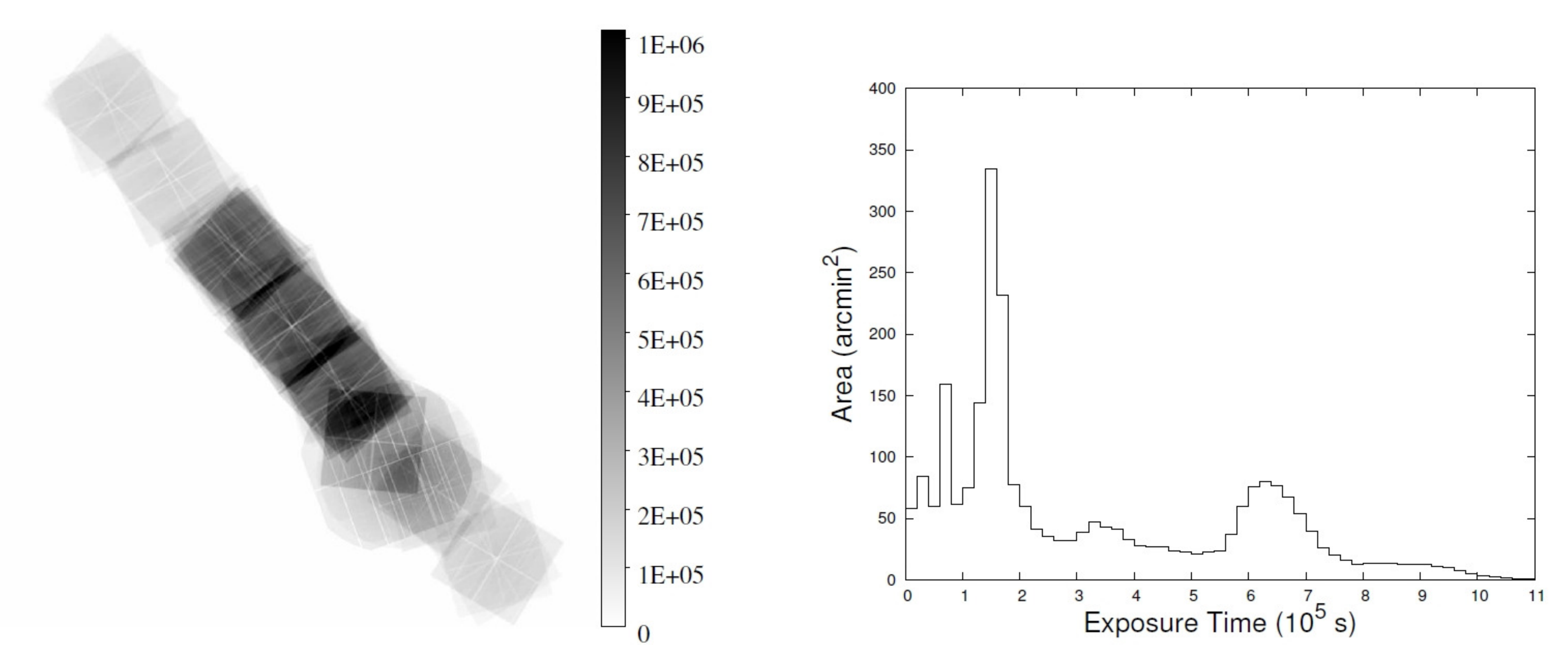}
\caption{The exposure map (left panel) and the distribution of exposure time (right panel) in Chandra and XMM coverage of EGS.\label{fig1}}
\end{figure*}

It is valuable to study galaxy groups over a range of cosmic time to understand the effect of the group
 environment on the galaxy population. Several efforts have been made to identify groups and clusters up to redshift
 one and beyond (e.g., \citealt{Stan06,Eis08,Bie10,Tan10}). The faint X-ray
 emission and low galaxy number densities of galaxy groups make such environments difficult to distinguish from the field compared to
 massive galaxy clusters at higher redshifts. We have a relatively good knowledge
 and samples of galaxy groups in the local universe (e.g. \citealt{Mul98}) but there is a lack of similar samples of galaxy groups
 which have both sufficiently deep X-ray data and counterparts with optical spectroscopy at
 high redshift. In the presence of advances in deep X-ray surveys, extended X-ray emission
 provides a reliable signal to detect such environments at high redshifts. 

 There are a number of different methods for detecting groups of galaxies:
 searches in optical data via the red-sequence method (e.g. \citealt{Gla05,Koe07});
 the Sunyaev-Zeldovich (SZ) effect on the cosmic microwave background, CMB
 (e.g. \citealt{Sun70,Sun72,Carl02,Lar03,Ben04,Stan09}; X-ray emission from hot intracluster gas
 (e.g. \citealt{Boe00,Has01,Vik09,Fin10}); cosmic shear due to weak gravitational lensing maps (e.g. \citealt{Miy07,Mas07});
 and spectroscopic group samples (e.g. \citealt{Ger12,Mil05,Kno09}).
 While spectroscopic surveys reveal the largest and deepest group
catalogs, detection of the group X-ray emission has been proven to
ensure objects are virialized, and with the deepest X-ray survey
available to date, the limits to which X-ray emission can
be detected are reaching the level of low-mass groups. Moreover, compared to shear maps, 
X-rays probe a wider range in mass and redshift (\citealt{Lea10}).

  In this paper, with recent deep Chandra data (Nandra et al. in prep), we search for galaxy groups in a wide
 range of redshift in the Extended Groth Strip (EGS). In previous work, searching for galaxy groups
 in Chandra data with a nominal exposure time of 200 ks and $\sim0.67  deg^2$ coverage of EGS field yielded a
 discovery of seven high significance galaxy groups in this field (\citealt{Jel09}). We now add
 new Chandra data with approximately 800 ks exposure time covering $\sim0.25 deg^2$ field; see Figure \ref{fig1}. Furthermore, EGS is one of the
 four fields in Deep Extragalactic Evolutionary Probe 2 (DEEP2) spectroscopy survey (\citealt{Dav03,New12}) and
 it is the only field which has been targeted for extensive spectroscopic data in DEEP3 (\citealt{Coo11,Coo12}).
 The DEEP2 and DEEP3 coverage of EGS are magnitude limited but not color selected, yielding a large sample of spectroscopic galaxies
 at all redshifts enabling us to identify our groups optically and determine their velocity dispersions.

 This paper is laid out as follows: section 2 presents a brief description of AEGIS survey and our data, section 3 describes our method for group identification.
 We present our identified group catalog in section 4. In section 5 we present spectroscopic group membership and dynamical properties of th groups. 
We make a comparison between the X-ray groups and optical groups which are
 identified from Voronoi-Delaunay method (VDM) in the DEEP2 spectroscopic dataset in section 6.$\;$Throughout this paper a $\Lambda CDM$
 cosmology with $\Omega_m=0.27$, $\Omega_\Lambda=0.73$ and $H_0=100 h$ km s$^{-1}$ Mpc$^{-1}$ where h=0.71 is assumed.\\

\section{THE AEGIS SURVEY}

The All-Wavelength Extended Groth Strip International Survey (AEGIS) brings together deep 
imaging data from X-ray to radio wavelengths and optical spectroscopy over a large area 
(0.5-1 $deg^2$). This survey includes: Chandra/ACIS X-ray (0.5-10 keV), GALEX ultraviolet 
(1200-2500 $\AA$), CFHT/MegaCam Legacy Survey optical (3600-9000 $\AA$), CFHT/CFH12K optical 
(4500-9000 $\AA$), Hubble Space Telescope/ACS optical (4400-8500 $\AA$), Palomar/WIRC near-infrared 
(1.2-2.2 $\mu m$), Spitzer/IRAC mid-infrared (3.6, 4.5, 5.8, 8 $\mu m$, VLA radio continuum 
(6-20cm) and a large spectroscopic dataset using the DEIMOS spectrograph on the Keck II 10m 
telescope in an area with low extinction and low Galactic and zodiacal infrared emission 
(Davis et al. 2007). In the following section we will describe the various data sets used in this analysis.

\subsection{X-ray data}
Our very deep Chandra survey used the Advanced CCD Imaging Spectrometer (ACIS-I) in three 
contiguous fields covering a total area of 0.25 $deg^2$ (Nandra et al. in prep)  and a 
series of eight pointings covering a total area of approximately $0.67 deg^2$ in the 
Extended Groth Strip. \citet{Lai09} provided the details for the latter survey 
and the X-ray point sources catalog. The total exposure time is approximately 3.4 Ms
 with nominal exposure of 800 ks in each three central fields. 

We also used the XMM-Newton observations of the field (ObjIDs 0127921001, 0127921101, 
0127921201, 0503960101), which were processed and co-added following the prescription 
of \citet{Bie10}. The total time of XMM observations is 100ks and its contribution 
to the final coverage can be seen in Figure \ref{fig1} as a roundish area in the southern part of 
the survey. Here we present the X-ray extended sources catalog based on the Chandra and XMM observations in EGS. 

In the Chandra analysis, we have applied a conservative event screening and modeling of the
quiescent background. We have filtered the light-curve events using the lc clean tool in 
order to remove normally undetected particle flares. The background model maps have been evaluated with the prescription
of \citet{Hic06}. We estimated the particle background by using the ACIS stowed position \footnote{http://cxc.cfa.harvard.edu/contrib/maxim/acisbg} observations and rescaling them by
 the ratio of the hard band (9.5$-$12keV) fluxes. The cosmic background flux has
been evaluated by subtracting the particle background maps from the real data and masking the area occupied
by the detected sources. We applied the method which has been used in \citet{Fin09} to search for extended sources and as a result we found 56 extended sources in EGS strip.
 Briefly, X-ray data have been obtained from X-ray mosaics made from coaddition of the $XMM-Newton$ and $Chandra$ data. After background subtraction
 and point source removal for each observation and each instrument separately, the residual images were
 co-added, taking into account the difference in the sensitivity of each instrument to produce a joint exposure map.
 To detect the sources we run a wavelet detection at $32''$ and $64''$ spatial scales,
 similar to the procedure described in \citet{Fin07,Fin09}.\\

\subsection{Photometric Data}\label{sec:data}
The EGS field is located at the center of the third wide field of the Canada-France-Hawaii Telescope
 Legacy Survey (CFHTLS-Wide3, W3)\footnote{Based on observations obtained with MegaPrime/MegaCam,
 a joint project of CFHT and CEA/DAPNIA, at the Canada-France-Hawaii Telescope (CFHT) which is
 operated by the National Research Council (NRC) of Canada, the Institut National des Science
 de l'Univers of the Centre National de la Recherche Scientifique (CNRS) of France, and the
 University of Hawaii. This work is based in part on data products produced at TERAPIX and
 the Canadian Astronomy Data Centre as part of the Canada-France-Hawaii Telescope Legacy Survey,
 a collaborative project of NRC and CNRS} which is covered in $u^\ast$, $g^\prime$, $r^\prime$,
 $i^\prime$ and $z^\prime$ filters down to $i^\prime$=24.5 with photometric data for 366,190 galaxies (\citealt{Bri08}). The EGS field also contains
 the CFHTLS Deep 3 field (\citealt{Dav07}), which covers 1 deg$^2$ with $ugriz$ imaging to depths ranging from 25.0 in $z$ to 27 in $g$.
 For this work, we have used the T0006 release of the CHTLS Deep data \footnote{http://terapix.iap.fr/cplt/T0006-doc.pdf}.
 The CFHTLS Deep field also contains near-infrared coverage in the $JHK$ bands via the WIRCam Deep Survey (WIRDS - \citealt{Bie12}).
 This covers 0.4 deg$^2$ of the D3 field and provides deep imaging to $\sim24.5$ (AB) in the three NIR bands. Photometric redshifts in the
 region covered by the NIR data were determined using the \texttt{Le Phare} code as described in \citet{Bie10}.

\subsection{Spectroscopic Data}
\label{sec:optfilt}
The DEEP2 Redshift Survey has targeted $\sim3.5$ $deg^2$ within four fields on the sky using the DEIMOS multi-object
 spectrograph (\citealt{Fab03}) on the Keck II Telescope (\citealt{Dav03}). All the DEEP2 targets have $18.5\leq R \leq24.1$. The EGS is one of these four
 fields. Compared to other DEEP2 fields, the EGS spectroscopy is magnitude limited, but not color-selected, giving the advantage of a sample of galaxies at
 all redshifts (\citealt{Dav07}). In addition, this region of sky has been targeted for extensive spectroscopy with DEEP3 (\citealt{Coo11,Coo12}). The DEEP2 and DEEP3
 catalogs have about 23,822 unique objects in total(with -2$\le$redshift quality$\le$4) and 16,857 objects with reliable redshifts (with redshift quality$\ge$3). 
In addition to DEEP2 and DEEP3, EGS is located in Sloan Digital Sky Survey \footnote{Funding for the SDSS and SDSS-II has been provided by the Alfred P. 
Sloan Foundation, the Participating Institutions, the National Science Foundation, the U.S. Department of Energy, the National Aeronautics and Space Administration, 
the Japanese Monbukagakusho, the Max Planck Society, and the Higher Education Funding Council for England. The SDSS Web Site is \url{http://www.sdss.org/}.} 
coverage so we have additional spectra for our low redshift galaxies. We also used redshifts of spectroscopic galaxies obtained in follow-up observations of 
the DEEP2 sample with the Hectospec spectrograph on the Multiple Mirror Telescope (MMT; \citealt{Coil09}).
\section{Optical Identification}
\label{sec:matchedfilt}
 To identify the groups in redshift space, we used galaxies with good redshift quality in DEEP2 and DEEP3
 to construct our initial redshift catalog. Using this catalog, imaging data, and position of X-ray extended sources, we assigned a redshift to each
 X-ray source visually where the spatial distribution of galaxies in the sky coincide with the X-ray emission. For those sources for which we found more than
 one counterpart and the X-ray shape allows to securely seprate the contribution from several counterparts, we define a new ID in our X-ray catalog. The detected sources areas
 searched for flux extension down to 90\% confidence level which is subsequently used for flux estimation.

 We also used the refined red-sequence technique, described in \citealt{Fin10}, to confirm the overdensity of red
 galaxies within or near those X-ray sources which have a lack of spectroscopic data. In brief, we selected galaxies with $|z-z_{phot}|<0.2$ and
 within a physical distance of 0.5$\,$Mpc from the center of X-ray emission at the given redshift. Then, using a Gaussian weight, we count galaxies around
 the model red-sequence and find overdensities. Since different observed colors are sensitive to red galaxies
 at different redshifts, we adopt the following combination of colors and magnitudes.\\
\\
For those groups which lie in D3 field in CFHTLS :\\
0.0$<$z$<$0.3  : $u^\ast-r^\prime$ color and $r^\prime$ magnitude \\
0.3$<$z$<$0.6  : $g^\prime-i^\prime$ color and $i^\prime$ magnitude \\
0.6$<$z$<$1.0  :  $r^\prime-z^\prime$ color and $z^\prime$ magnitude \\
1.0$<$z$<$1.5  : $i^\prime-J$ color and J magnitude \\
1.5$<$z$<$2.0  : $z^\prime-Ks$ color and Ks magnitude \\
\\
For those in the W3 field:\\
 0.0$<$z$<$0.3 : $u^\ast-r^\prime$ color and $r^\prime$ magnitude\\
 0.3$<$z$<$0.6 : $g^\prime-i^\prime$ color and $i^\prime$ magnitude\\
 0.6$<$z : $r^\prime-z^\prime$ color and $z^\prime$ magnitude

We note that \citet{Bie10} mistakenly quoted their filter combinations to identify red-sequence signals. They used the same 
filters as in our D3 field. Using the red-sequence technique and the spectroscopic data, we could identify redshifts for 52 extended 
X-ray sources. We have spectroscopic redshifts for 49 galaxy groups and more than two spectroscopic members for 46 galaxy groups. 

 We also assigned a flag for each extended source that describes the quality of the identification. Flag=1 indicates confident redshift assignment and significant X-ray emission and also good
 centering, while for Flag=2 the centering has a large uncertainty. In the cases when a single X-ray source has been matched to several optical counterparts
 the assigned flag is equal to 2 or larger. For Flag=3 we have no spectroscopic confirmation but good centering and for Flag=4 we have unlikely redshifts
 due to the lack of spectroscopic objects and red galaxies and also a large uncertainty in centering. We assigned Flag=5 for the 13
 unreliable cases for which we could not identify any redshift. They can be split into the following categories.

 Some of the X-ray extended sources do not have spherical and symmetric morphologies and some exhibit a secondary peak in X-ray distribution.
 Initially, we expect this results from overlapping systems but visual inspection of optical data and the red-sequence method indicate a single significant group for some of these. So, 
we classify the second shallow peak in X-rays as substructure inside those real groups. Furthermore, sources on the edge of X-ray coverage with low signal to noise and no optical
 counterparts could be explained as residual background level in the images or bright X-ray clusters outside the field of view. We note that two RCS-2 clusters (\citealt{Gil11})
 are located $10'$ west from the edge of the survey. In the few cases of bright stars near the X-ray emission, we could not match galaxy counterparts to the extended emission.
 We assigned Flag=5 for all of these cases and they are not included in the final sample.

 \begin{figure}[t]
 \includegraphics[width=8cm]{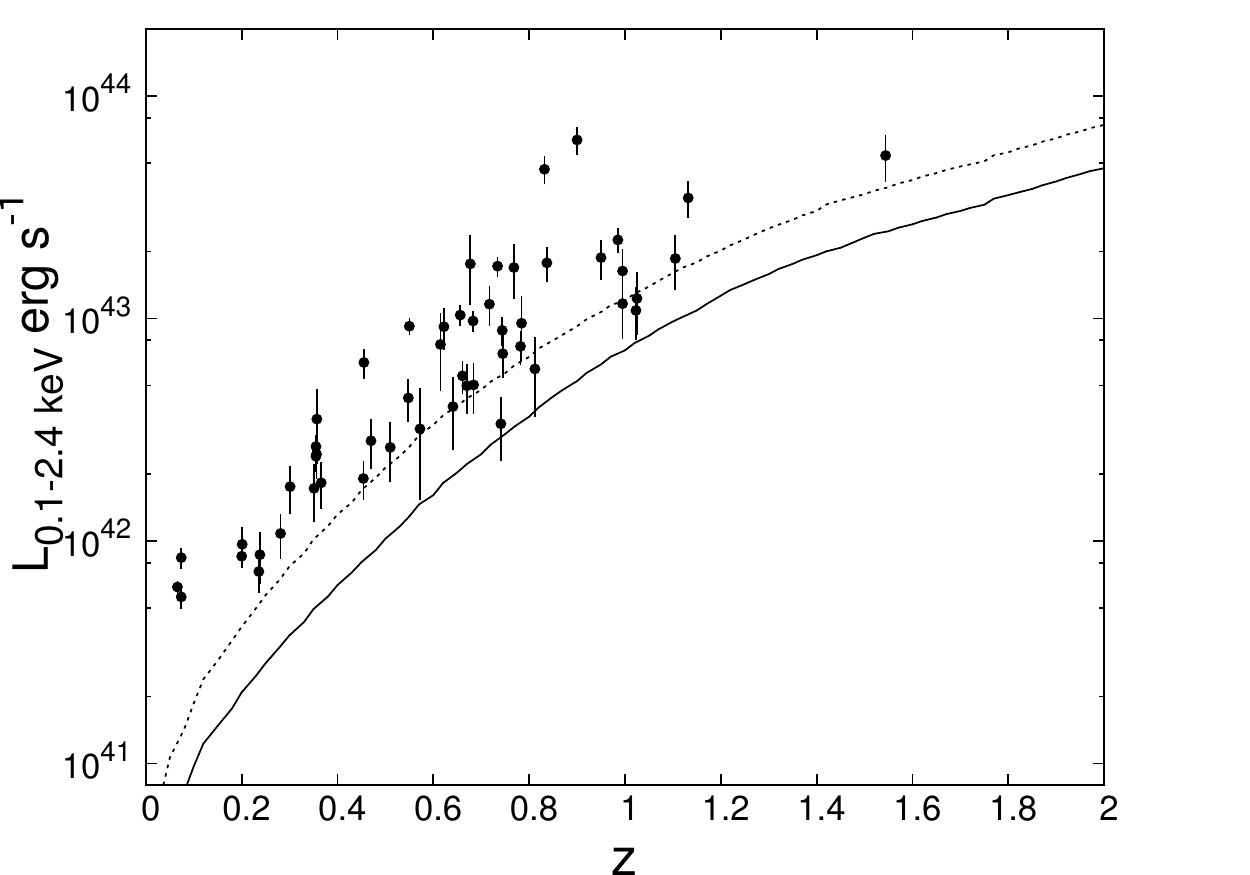}
 \figcaption{X-ray luminosity as a function of redshifts for X-ray galaxy groups in EGS. The error bars are based only on the statistical errors in the flux measurements. The solid line and dashed line are
 the flux detection limits associated with 10 and 50\% of the search area respectively.\label{fig2}}
 \end{figure}
\begin{figure}[t]
\includegraphics[width=8cm]{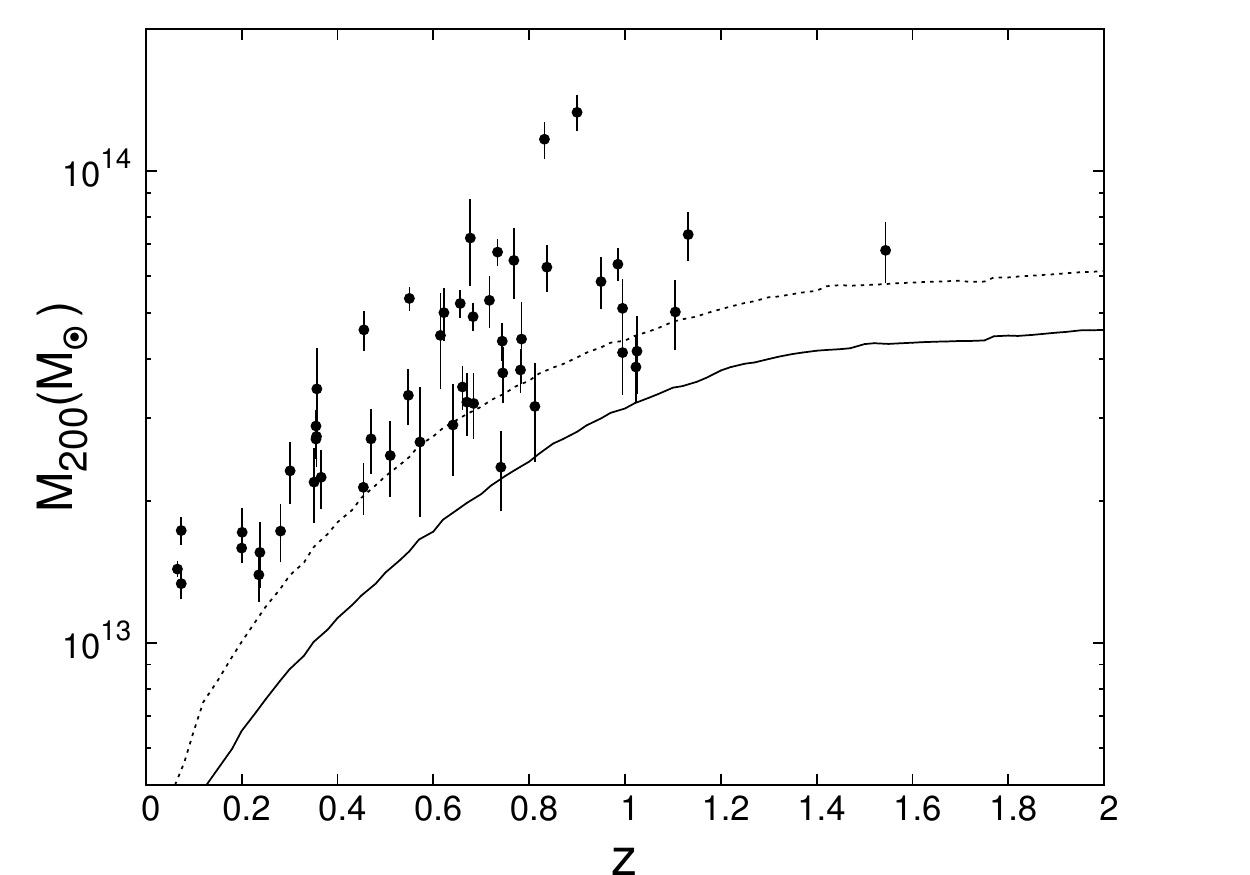}
\figcaption{X-ray masses as a function of redshifts for X-ray galaxy groups in EGS. The error bars are based only on the statistical errors in the flux measurements. 
The solid line and dashed line are the flux detection limits which associate 
with 10 and 50\% of search area respectively.\label{fig3}}
\end{figure}

\section{A Catalog of identified X-ray groups}
In this section, we describe our catalog of 52 X-ray galaxy groups detected in AEGIS (Table 1). The group identification number, RA and Dec. of the peak of X-ray emission
 in Equinox J2000.0 are listed in Column 1, 2 and 3. In Column 4 the mean of red-sequence redshifts which is substituted with the median of spectroscopic redshifts in
 case there is a spectroscopic redshift determination for the group member galaxies is listed. We provide the group flux in the 0.5--2 keV band in Column 5 with the corresponding
 1$\sigma$ error. The rest-frame luminosity in the 0.1--2.4 keV is given in Column 6. Column 7 lists the estimated total mass, $M_{200}$, computed following \citealt{Lea10} 
and assuming a standard evolution of scaling relations: $M_{200}E_z=f(L_xE^{-1}_z)$ where $E_z=(\Omega_M(1+z)^3+\Omega_\Lambda)^{1/2}$. The corresponding
 $r_{200}$, $M_{200}=\frac{4}{3}\pi r_{200}^3(200 \rho_{critical})$, in arcminutes is given in Column 8. Column 9
 lists the flag for our identification, as described in section 3. The number of spectroscopic member galaxies inside $r_{200}$ is given in Column 10 (see $\S5$). Column 11 lists flux 
 significance which provides insight on the reliability of both the source detection and the identification. The velocity dispersion estimated from X-ray luminosities is given in column 12.\\

Figures \ref{fig2} and \ref{fig3} show the luminosity and mass of groups as a function of their redshifts respectively.

\subsection{A galaxy group candidate at $z$=1.54}
\begin{figure}[t]
\centering
\includegraphics[width=6.5cm]{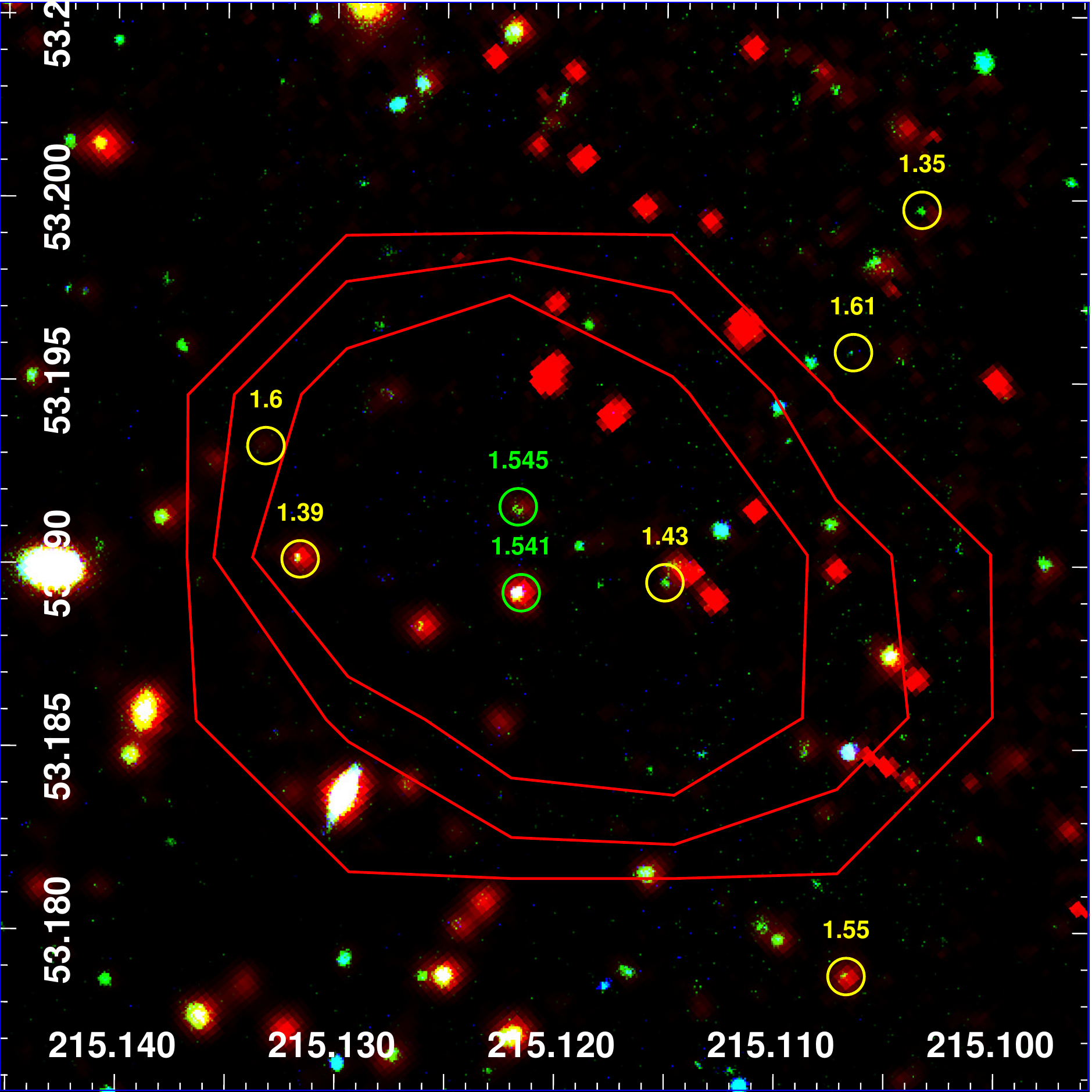}
 \figcaption{RGB image of galaxy group at z=1.54 with ID = EGSXG J1420.4+5311 using channel 2 (4.5 microns) of Spitzer/IRAC and $z^\prime$ and $r^\prime$ bands from W3 field
 of CFHTLS. The contours show the X-ray emission. The green circles show spectroscopic redshifts and the yellow circles indicate galaxies
 located at $1.3<z_{photo}<1.7$ within $r_{200}$ of the group. Many of the red points are artifacts in the ch2 image and they don't have any corresponding sources in $z^\prime$ and $r^\prime$ bands.\label{fig4}} 
 \end{figure}

\begin{figure}[t]
\centering
 \includegraphics[width=8cm]{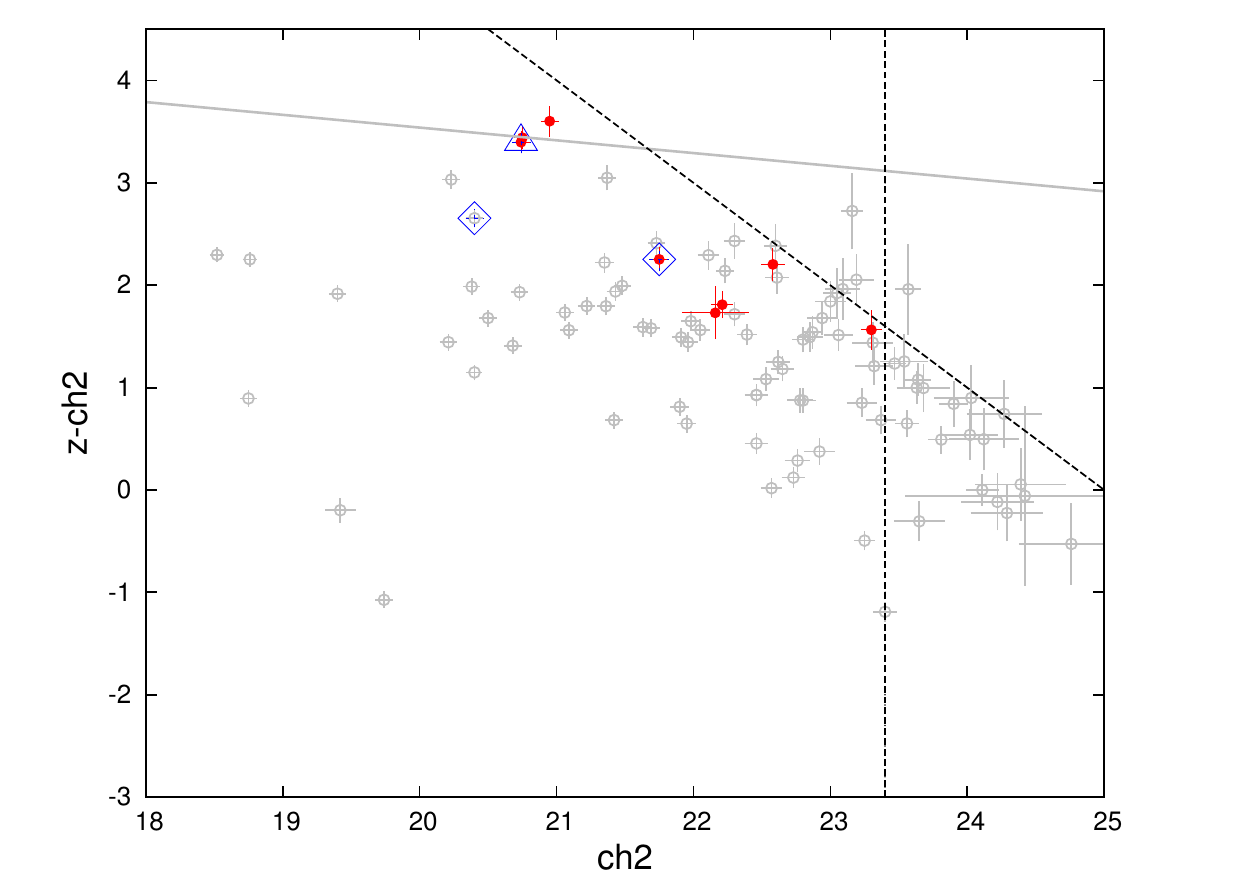}
 \figcaption{Color-magnitude diagram for EGSXG J1420.4+5311 based on Spitzer/IRAC and CFHTLS data. All the galaxies within $r_{200}$ of the group are plotted here.
 The filled circles show galaxies with $1.3<z_{photo}<1.7$. The diamonds and triangle indicate secure and possible spectroscopic members respectively. The grey line shows model red-sequence for z=1.5. The 50\%
completeness for Channel 2 magnitude and z-ch2 color are shown as vertical and slanted dashed lines respectively.\label{fig5}} 
\end{figure}

During the optical group identification using spectroscopic data, we have discovered a high-z group candidate at z=1.54 (Figure \ref{fig4}). 
The X-ray signal is measured with a significance of $~4.1\sigma$. This group has two sepectroscopic members with good flags in the hot halo of the group.
 One of the spectroscopic members show AGN activity in its expectra and also detected as a X-ray point source in our analysis. 
The point source emission has been removed from the flux estimates. As it is a Chandra-detected group, the resolution allows us to exclude
 the AGN contamination down to a factor of 10 below the level of the detection of the extended emission itself. We estimate a cluster mass of $M_{200}= 6.8 \times 10^{13} M_\odot$, an X-ray
 luminosity of $L_X= 5.4 \times 10^{43}$ erg s$^{-1}$ and a virial radius of $r_{200}=0.015^\circ$. A red-sequence finder using channel 2 (4.5 micron) from Spitzer/IRAC and $z^\prime$ band from CFHTLS has
 also detected a signal around z=1.5 (Figure \ref{fig5}). Since we did not have a deep z-band image for this group, there was a strong limit on our color-magnitude diagram and thus the
 red-sequence signal and therefore we called it a group candidate. The uniqueness of this candidate group arises from availability of ultra deep X-ray image but the system is marginally
 covered by Spitzer/IRAC data and is out of coverage of Deep fields in AEGIS (CFHTLS D3, Hubble/ACS and CANDELS). 

\section{Spectroscopic group member galaxies}
We search for galaxies associated with our identified X-ray sources based on their redshifts and positions. We perform this selection in different ways and explore the effects on the dynamical
 velocity dispersion, mass, and $L_x-\sigma$ scaling relation of the groups.

 First, we assume an initial velocity dispersion of 500 km/s for each group and calculate the redshift range for the group members from equation 1 (\citealt{Wil05,Con12}):
\begin{equation}
 \delta(z)_{max}=2\frac{\sigma(v)_{obs}}{c}
\end{equation}
This $\delta(z)_{max}$ is then converted into a spatial distance using equation 2 and 3:

\begin{equation}
 \delta(r)_{max}=\frac{c\delta(z)_{max}}{b \cdot H_{71}(z)}
\end{equation}
with $b=AspectRatio=9.5$\\
\begin{equation}
 \delta(\theta)_{max}=206265''\frac{ \delta(r)_{max}}{h^{-1}_{71}Mpc} \cdot (\frac{D_{\theta}}{h^{-1}_{71}Mpc})^{-1}
\end{equation}
where $D_{\theta}$ is angular diameter distance.\\
Considering the center of the X-ray emission as the center of the groups, we selected groups members which lie within our redshift and angular limits (equation 4 and equation 5). 
\begin{equation}
|z-z_{group}|<\delta(z)_{max}
\end{equation}
\begin{equation}
\delta(\theta)<\delta(\theta)_{max}
\end{equation}
We recompute the observed velocity dispersion of the groups, $\sigma(v)_{obs}$ using the ``gapper'' estimator method which gives more accurate measurement of velocity dispersion for
 small size groups (\citealt{Bee90,Wil05}) in  comparison to the usual formula for standard deviation,
 $\sigma^2=\langle v^2 \rangle-\langle v \rangle^2$.
 According to the formula\\
\begin{equation}
\label{1}
\sigma(v)_{obs}=1.135c\times\frac{\sqrt\pi}{N(N-1)}\sum_{i=1}^{n-1}\omega_ig_i
\end{equation}
where $w_i=i(N-i)$ , $g_i=z_{i+1}-z_i$ and $N$ is the total number of spectroscopic members. In this way we measure the velocity dispersion using the line-of-sight velocity 
gaps where the velocities have been sorted into ascending order. The factor 1.135 corrects for the 2$\sigma$ clipping of the Gaussian velocity distribution.
We then consider the r-band luminosity-weighted centroid in projected space as the center of the group and the mean redshift of the galaxy members as the group redshifts and again find
 the galaxy members. We repeat the entire process until we obtain a stable membership solution. For all the groups, we reach such a stable membership after 2 iterations. At the end, 
 we calculated the rest-frame and intrinsic velocity dispersion according to
\begin{equation}
\label{eqn:1}
\sigma(v)_{rest}=\frac{\sigma(v)_{obs}}{1+z}
\end{equation}
\begin{equation}
\label{eqn:2}
\langle\Delta(v)\rangle^2=\frac{1}{N}\sum_{i=1}^N\Delta({v})_i^2
\end{equation}
\begin{equation}
\label{eqn:3}
\sigma(v)_{intr}^2=\sigma(v)_{rest}^2-\langle\Delta(v)\rangle^2
\end{equation}
The intrinsic velocity dispersion, $\sigma(v)_{intr}$, is computed by removing the effect of measurement errors of component galaxies from the rest-frame
 velocity dispersion, $\sigma(v)_{rest}$ (equations 8 and 9). Then we calculated errors for our velocity dispersions using the Jackknife technique (\citealt{Efr82}).
 The error is $[\frac{N}{N-1}\sum(\delta_i^2)]^\frac{1}{2}$ where $\delta_i=\sigma(v)_{obs}-\sigma(v)_{obs,excluding\, i_{th}\, member}$.\\
We then applied two different optical and X-ray based cuts for the radius used to select member galaxies. Using $\sigma(v)_{intrinsic}$ derived from member galaxies after iterating,
 we computed $r_{200}$ for the optical cut as \citet{Carl97}:
 \begin{equation}
 \label{eqn:4}
 r_{200}=\frac{\sqrt3 \sigma_{intr}} {10H(z)}
\end{equation}
 Where
\begin{equation}
 H(z)=H_0E(z)
\end{equation}

\begin{figure*}
\centering
 \includegraphics[scale=0.4]{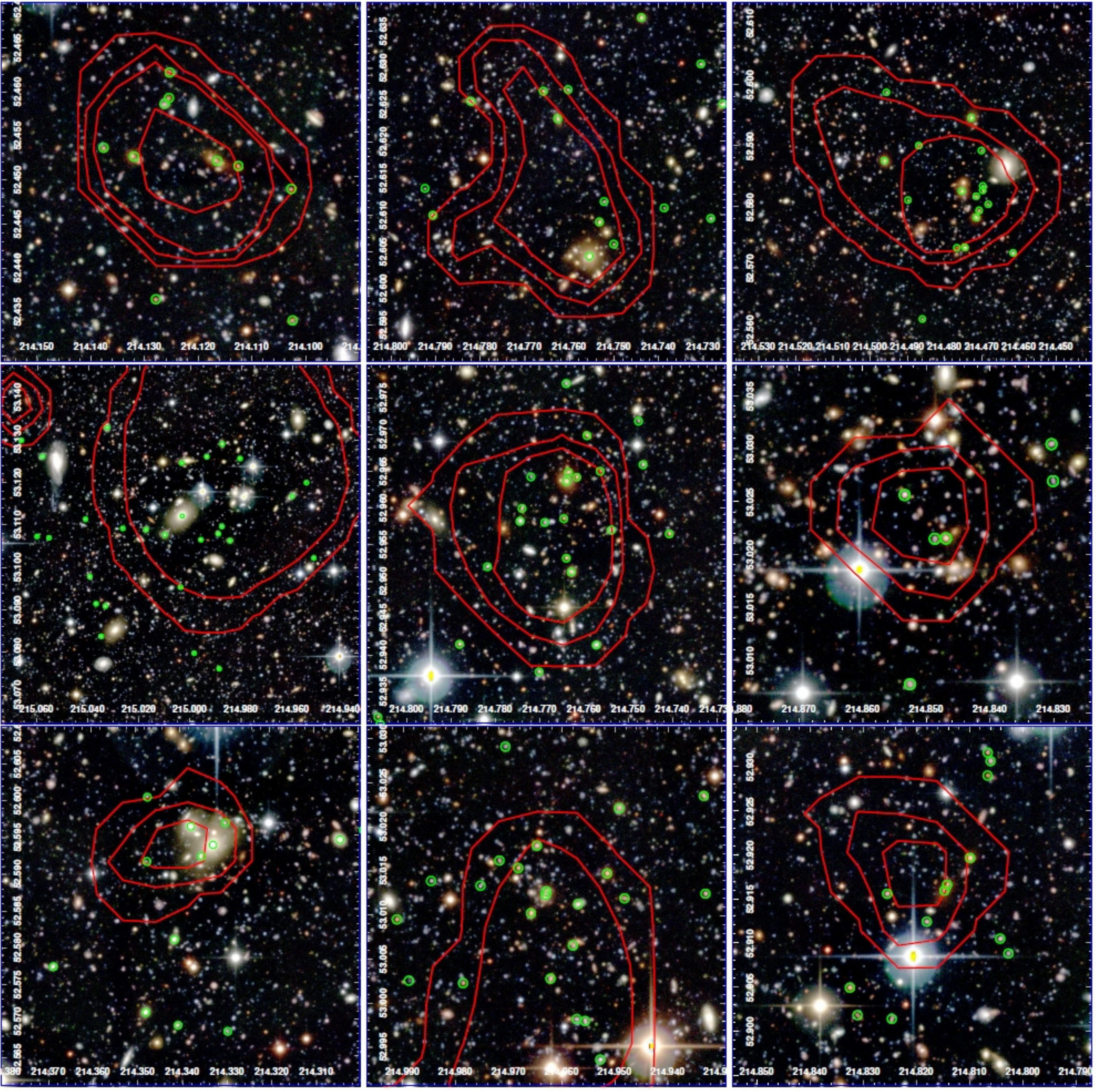}
\figcaption{CFHTLS D3 RGB images of X-ray galaxy groups using  $g^\prime$, $r^\prime$ and $i^\prime$ bands of the groups with
contours indicating levels of X-ray emission (in red) and spectroscopic members of the groups inside the virial radius estimated from X-rays. From upper left to 
lower right, the group IDs are: EGSXG J1416.4+5227,  EGSXG J1419.0+5236, EGSXG J1417.9+5235, EGSXG J1420.0+5306, EGSXG J1419.0+5257, EGSXG J1419.4+5301, 
EGSXG J1417.3+5235, EGSXG J1419.8+5300, and EGSXG J1419.2+5255. The horizontal and vertical axes show the right ascension and declination respectively.\label{fig6}}

\end{figure*}

For the X-ray defined radial cut, we used $r_{200,x}$ from the main catalog(see $\S$ 4). We then calculated the intrinsic velocity dispersion for all groups based on equations 6, 7, 8 and 9 for the 
members with the optical and X-ray based radial cuts. In Figure \ref{fig6} we present a gallery of RGB images of the X-ray galaxy groups within the D3 field of CFHTLS.

In the second way, we pick the spectroscopic galaxies with positions within the $r_{200,x}$ of the X-ray centers and their redshifts match 
to $\Delta z_1=0.001\times(1+z_{G})$ and $\Delta z_2=0.0025\times(1+z_{G})$ which corresponds to typical minimum and maximum velocity dispersions of a group. 
Table 2 shows a sample of spectroscopic galaxy members based on $0.0025\times(1+z_{G})$ selection. Then we computed intrinsic velocity dispersions for the member galaxies based on the 
X-ray virial radius and two different redshift cuts ($\Delta z_1$ \& $\Delta z_2$). 

\begin{figure}[t]
\includegraphics[width=8cm]{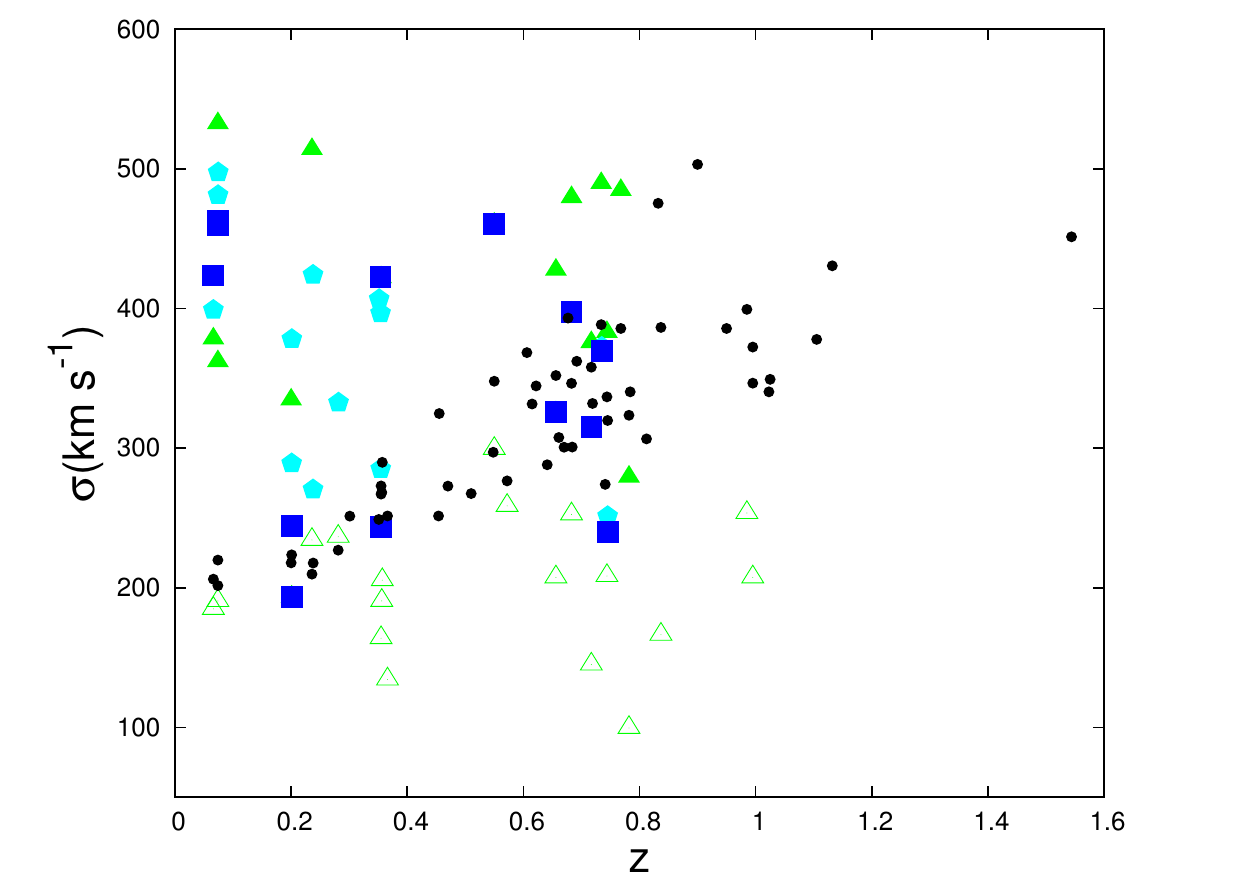}
\figcaption{Velocity dispersion as a function of redshift. The black filled circles show the velocity dispersions estimated from X-ray luminosities using scaling relations for all the groups. 
The filled and empty triangles show velocity dispersions for the groups with $\Delta z_2$ and $\Delta z_1$ respectively. The pentagons show the velocity dispersion 
for the groups with optically determined radial cut after iterations and the squares show the velocity dispersion for the groups with X-ray radius cut after iterations.\label{fig7}} 
 \end{figure}

In all cases, we considered galaxy groups with more than 10 members when we use the \textquotedblleft gapper\textquotedblright estimator to have a more relaible 
measurement of velocity dispersion(e.g. \citealt{Zab98,Gir01}). Figure \ref{fig7} shows velocity dispersions of X-ray galaxy groups derived by \textquotedblleft gapper\textquotedblright 
estimator method and from X-ray emission.

\subsection{The relation between X-ray luminosity and dynamical velocity dispersion}
\begin{figure}[t]
\includegraphics[width=8cm]{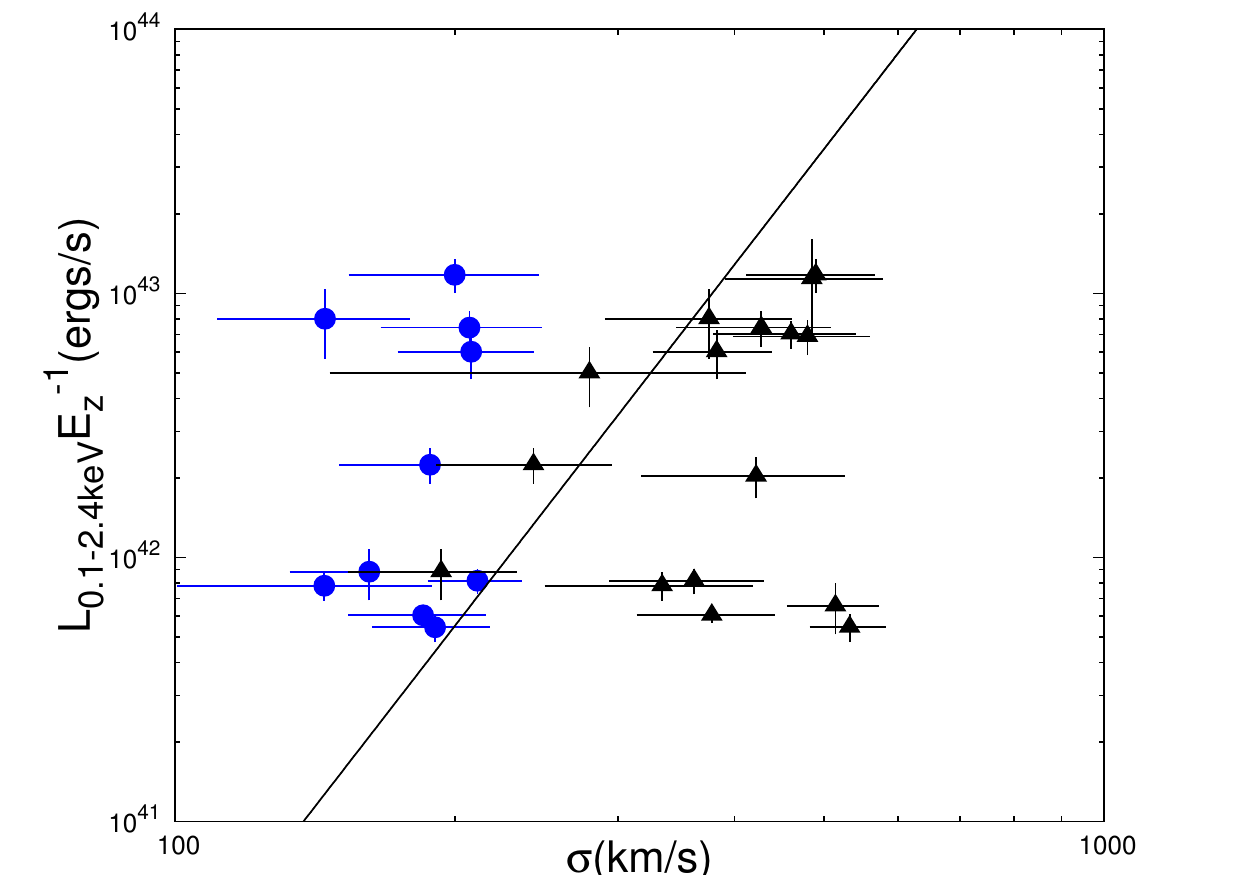}
 \figcaption{$L_X-\sigma$ relation for X-ray groups. The blue circles and the black triangles are corresponding to groups which members match to $\Delta z_1$ and $\Delta z_2$ respectively. 
The solid line show our expectation for $L_X-\sigma$ relation from scaling relations.\label{fig8}} 
\end{figure}

\begin{figure}[t]
 \includegraphics[width=8cm]{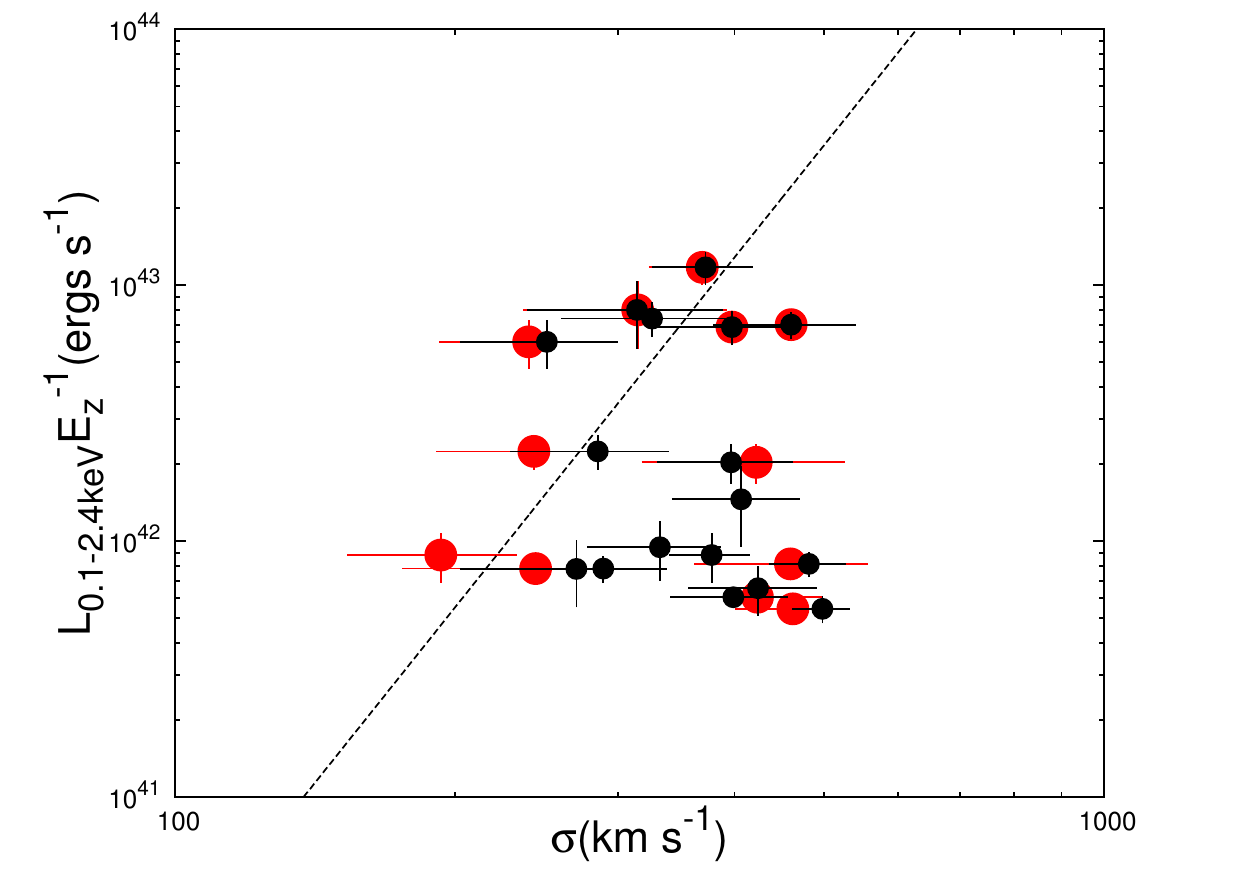}
 \figcaption{$L_X-\sigma$ relation for X-ray groups which members lie within a dynamically based virial radius (black circles) and X-ray based virial radius (red circles).\label{fig9}}
 \end{figure}

Figures \ref{fig8} and \ref{fig9} show the X-ray luminosity versus velocity dispersion for different methods. These plots include all the galaxy groups with Flag=1 and 2. We also plot
the $L_x-\sigma$ relation (dashed line) expected from scaling relations obtained for a sample of groups with similar luminosities in the $0<z<1$ redshift range in COSMOS (\citealt{Lea10}).
Velocity dispersions can be biased to higher values for low mass systems when we select members based on large $\Delta z$ as we may include outliers in our calculation and conversely can be biased to lower 
values for high mass systems if we select member galaxies within low $\Delta z$ along the line of sight as we are then ignoring some parts of the group (Figure \ref{fig8}). Furthermore, tracking the groups while we use 
different $\Delta z$ to choose member galaxies reveals an average systematics error $\sim 190$ km/s. \\
Figure \ref{fig9} shows the $L_x-\sigma$ relation for galaxy groups which members are selected based on two different radial cuts. It is obvious from Figure \ref{fig9} that different radial cuts can cause a change in
 scatter of $L_x-\sigma$ relation but, in the case of high mass systems, there is no apparent change in scatter of this relation. In addition, low X-ray luminosity systems show significant deviations
 from the scaling relation in both X-ray and optically based radial cuts. 

We looked at the quality flags, dynamical complexity and the X-ray compactness in comparison to the virial radius for the groups in order to study the group properties and their effects on the relation.

\begin{figure}[t]
 \includegraphics[width=8cm]{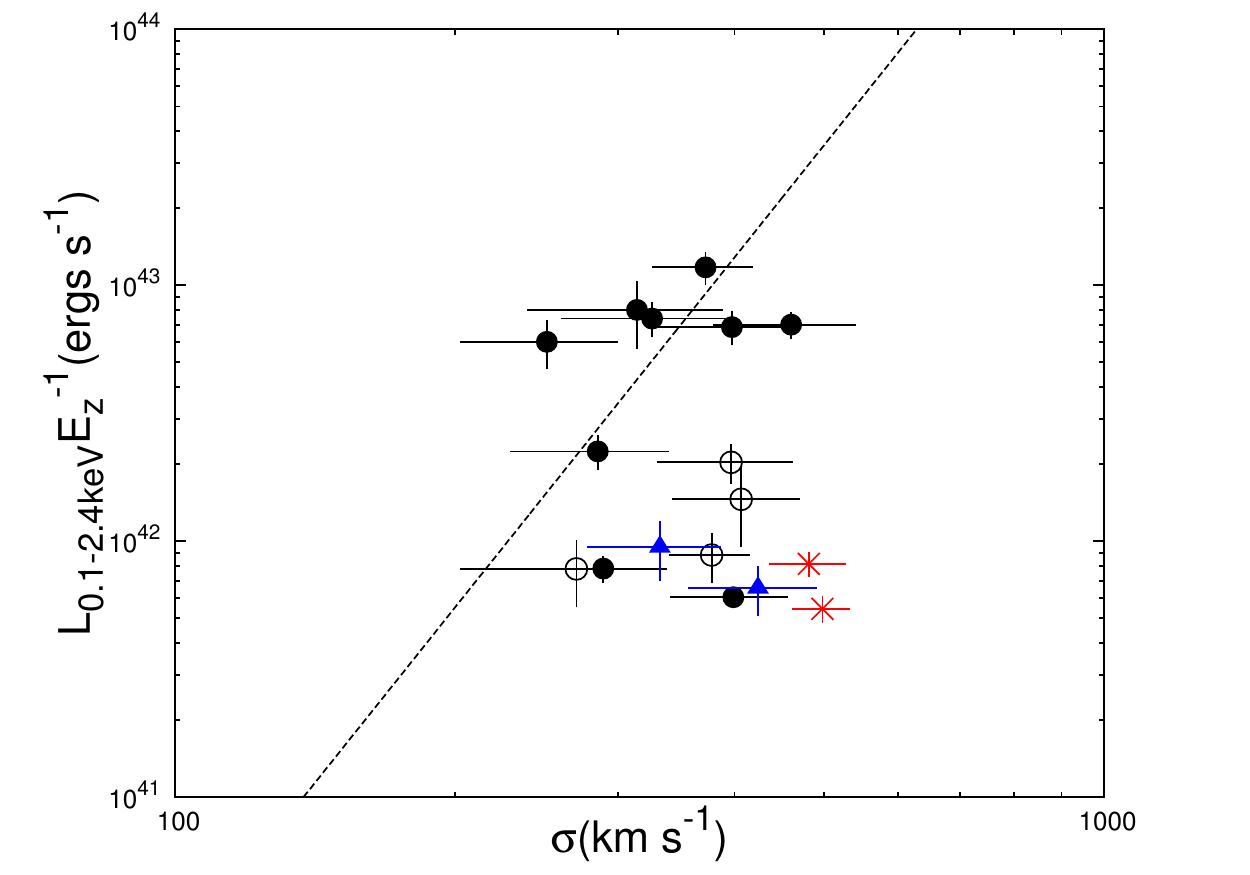}
 \figcaption{$L_X-\sigma$ relation for X-ray groups which members lie within a dynamically based virial radius. The red stars show groups which have substructure detected by DS test,
 blue triangles are the compact X-ray systems, and open circles show the groups with Flag=2. The solid line shows our expected $L_X-\sigma$ relation derived from scaling relations.\label{fig10}}
 \end{figure}

\begin{figure}[t]
 \includegraphics[width=8cm]{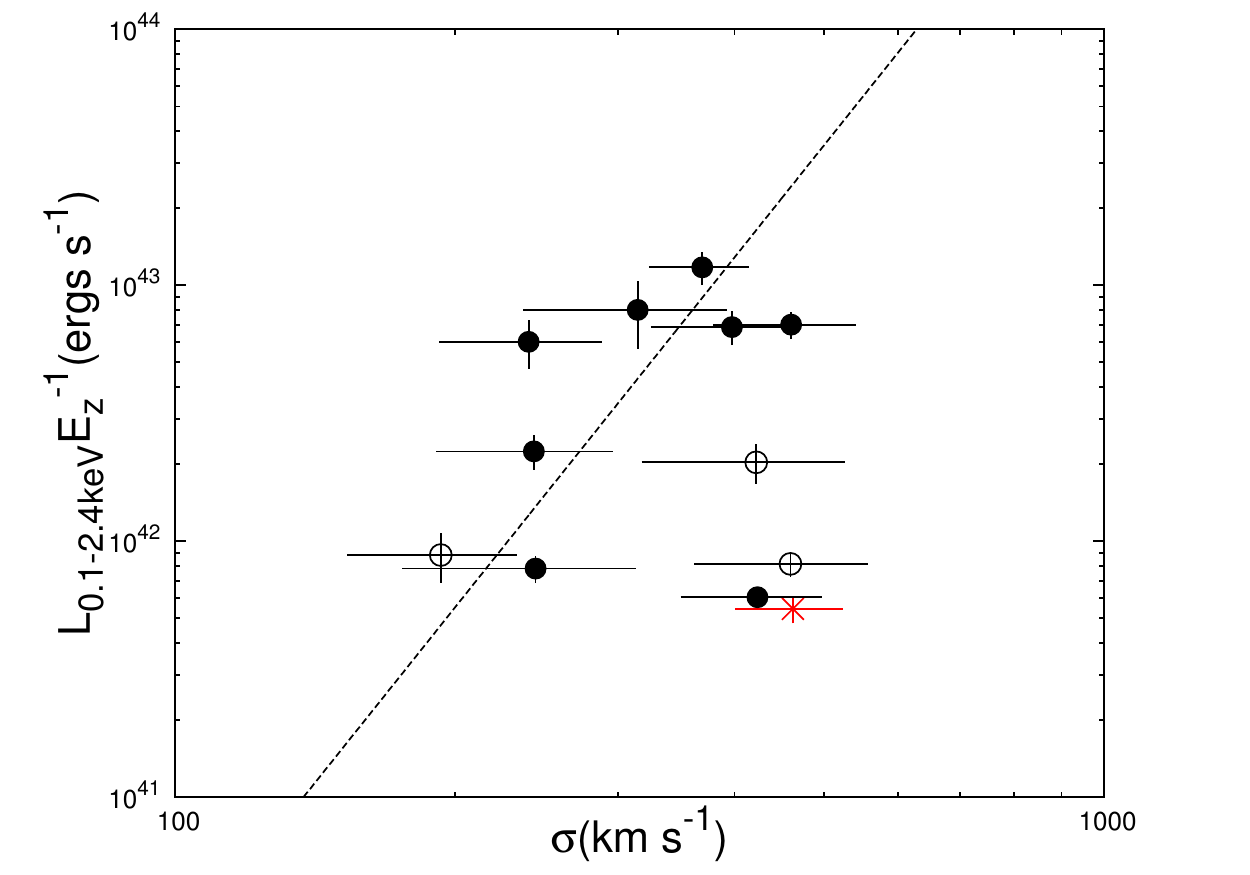}
 \figcaption{$L_X-\sigma$ relation for X-ray groups given a raidal cut based on X-ray. The red stars show groups which have substructure detected by DS test,
 blue triangles are the compact X-ray systems and open circles show the groups with Flag=2. The solid line show our expectation 
for $L_X-\sigma$ relation from scaling relations.\label{fig11}}
 \end{figure}

 As we expect, Figures \ref{fig10} and \ref{fig11} illustrate galaxy groups with Flag=2 have significant deviations from the relation compared to galaxy groups with Flag=1 (similar to what \citealt{Con12} found
 for the intermediate redshift X-ray selected groups).

To search for substructure in our groups, we apply the Dressler-Shectman (DS; \citealt{Dre88}). We use the DS test as in \citet{Hou12} which implement it 
for group size systems. In brief, we consider each individual galaxy in the group plus the $N_{nn}$ nearest members to it with $N_{nn}=\sqrt n_{mem}$ 
and calculate mean velocity and velocity dispersion for them ($\bar{v}^i_{local},\sigma^i_{local}$). Then we compute the deviations for each galaxy from the mean velocity ($\bar{v}$) and 
velocity dispersion ($\sigma$) of the whole group with $n_{mem}$ galaxies:
\begin{equation}
 \delta_i^2=(\frac{N_{nn}+1}{\sigma^2})[(\bar{v}^i_{local}-\bar{v})^2-(\sigma^i_{local}-\sigma)^2]
\end{equation}
where $1\le i \le n_{members} $.
 Then $\Delta$ statistics were computed using:
\begin{equation}
\Delta=\sum_{i=1} \delta_i
\end{equation}

To identify substructure, we used a probability (P-values) threshold for the DS test so we run 10,000 Monte Carlo simulations for each group. In each Monte Carlo run, the observed velocities are 
randomly shuffled and reassigned to member positions and $\Delta_{shuffled}$ is computed. The probabilities are given by
\begin{equation}
P=\sum(\Delta_{shuffled}>\Delta_{observed})/n_{shuffle}
\end{equation}
$n_{shuffle}$ is the number of the Monte carlo simulations which in our case is 10,000. A system is then considered to have significant substructure with 99 percent confidence level when $P < 0.01$. In total,
 we found 2 galaxy groups of 17 in optically based radial cut groups and 1 group of 12 in X-ray based radial cut galaxy groups with significant substructure which are marked with stars in Figure \ref{fig10} and \ref{fig11}.
\begin{figure*}[t]
\centering

 \includegraphics[width=15.cm]{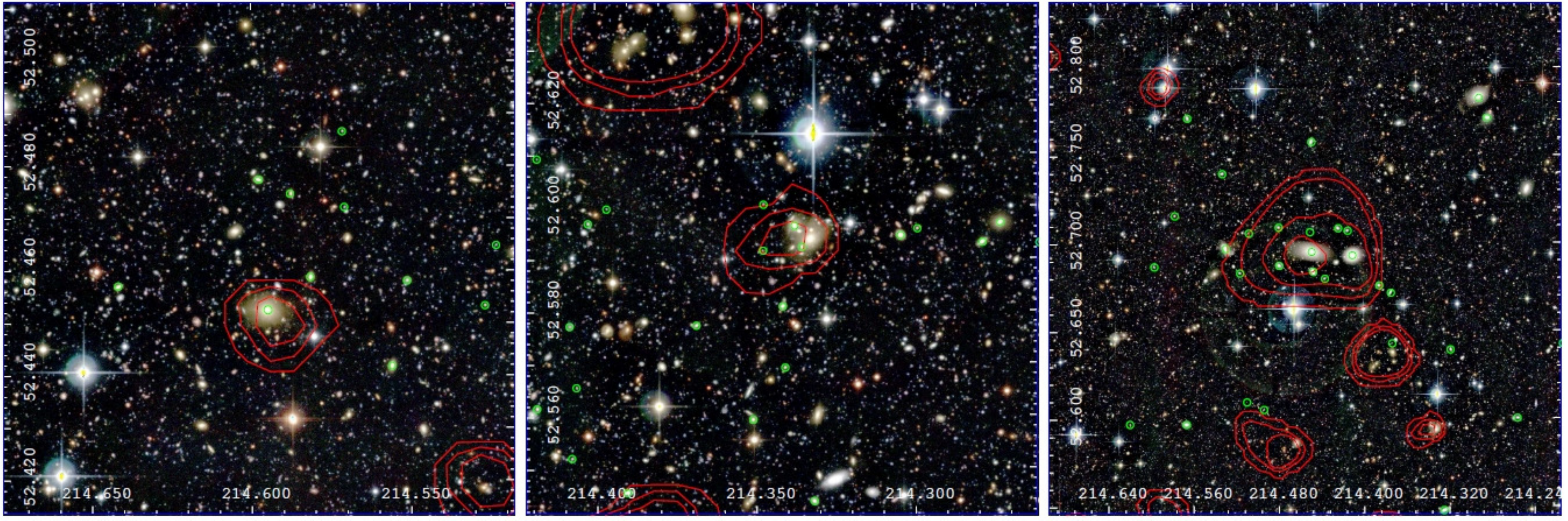}

\figcaption{CFHTLS D3 RGB images of X-ray galaxy groups using  $g^\prime$, $r^\prime$ and $i^\prime$ bands of the groups with
contours indicating levels of X-ray emission (in red) and spectroscopic members (in green). The groups ID from left to right are: EGSXG J1418.3+5227, EGSXG J1417.3+5235 and EGSXG J1417.7+5241\label{fig12}}

\end{figure*}
Figure \ref{fig12} shows the optical images for three galaxy groups which have the largest deviations from the scaling relation in Figure \ref{fig10}. All of them have Flag=1 and subtructure is not detected using the DS test.
The two left images in Figure \ref{fig12} have less than 10 members given an X-ray based virial radius cut and are not included in Figure \ref{fig11}. We compared the extension of X-ray emission to virial radius extracted from X-rays and optical 
velocity dispersion for galaxy groups. As Figure \ref{fig13} shows there is a population of the group galaxies which the fraction of the extension 
of X-rays to virial radius is less than 20\% in optically based radial cut and less than 15\% in X-ray based radial 
cut. These populations are dominated by some galaxy groups with Flag=2 and two galaxy groups from left in Figure \ref{fig12}  with an over-luminous galaxy close to the X-ray
 center.

\begin{figure}[t]
 \includegraphics[width=9cm]{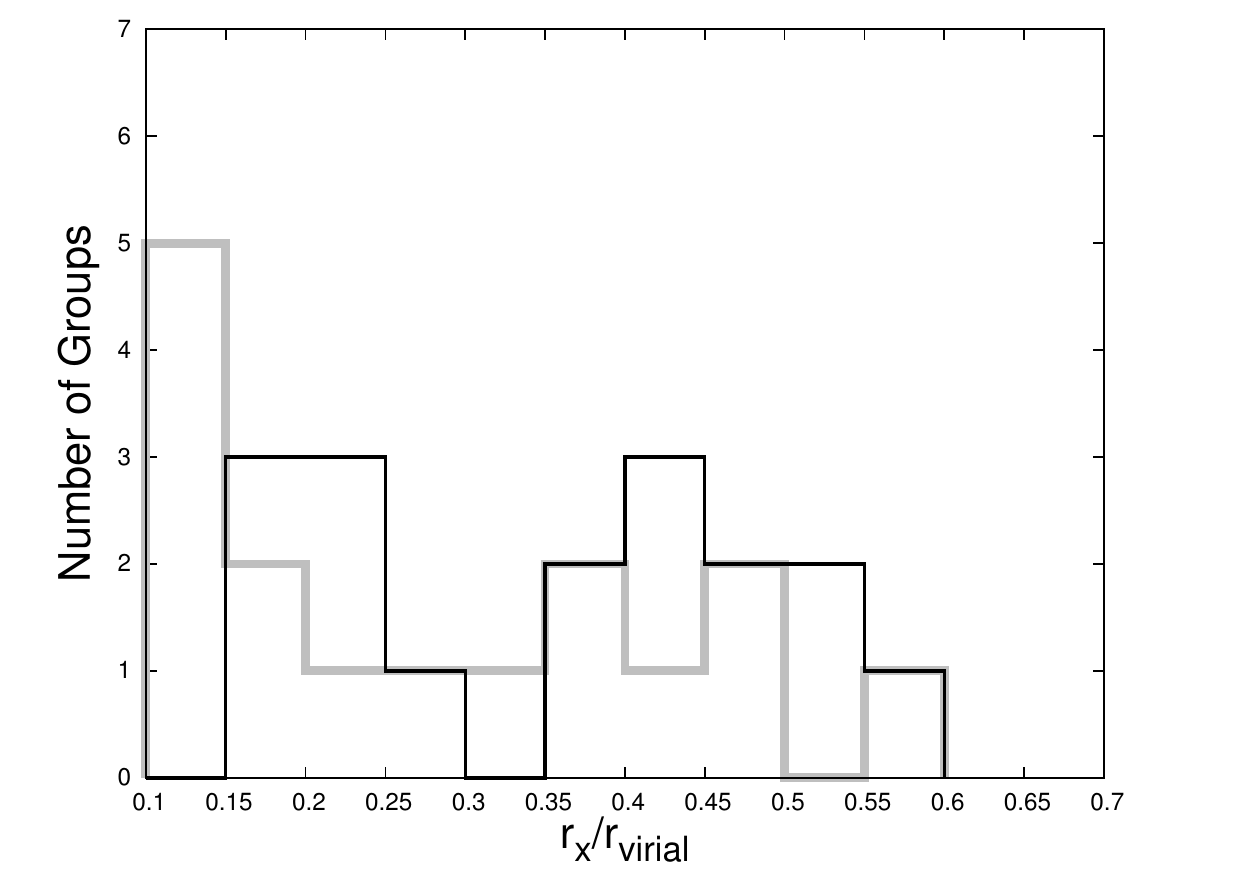}
 \figcaption{Fraction of X-ray extent to virial radius for an X-ray based virial radius ( black line) and optically based virial radius (grey line).\label{fig13}}
 \end{figure}

 We explored the $L_x-\sigma$ relation in more detail for three of the groups with large deviations from the scaling relation (Figure \ref{fig12}).
 EGSXG J1418.3+5227 with $L_x=1.08\times10^{42}$ erg s$^{-1}$, has $\Delta m_{12}=2$ where $\Delta m_{12}$ is the r-band magnitude difference between the first
 and the second brightest galaxy located in half of the virial radius of the group. These conditions result in  the classification of this group as a fossil galaxy group candidate (\citealt{Jon03}).
 As fossil galaxy groups are believed to be the final result of galaxy merging in normal groups, we expect a sufficiently deep potential well and high X-ray luminosity for these systems
 (\citealt{pon94, Jon00}). As a consequence, we expect fossil groups be more X-ray luminous than normal groups for a given velocity dispersion (\citealt{sha07})
 but instead we find the opposite. Moreover, \citet{Osm04} applied a radius of 60 kpc as a threshold for detectable X-ray emission to separate galactic haloes from group-scale haloes using
 different studies of bright isolated galaxies (\citealt{Osu03,Osu04}). The radius of detectable X-ray emission (at which the group
emission fell to the background level ) for EGSXG J1418.3+5227 is more than this threshold and about 95 kpc. However, the nature of this X-ray extended source with such unexpectedly low X-ray emission 
compared to the velocity dispersion is still a matter of interest.

\begin{figure}[t]

 \includegraphics[width=8.cm]{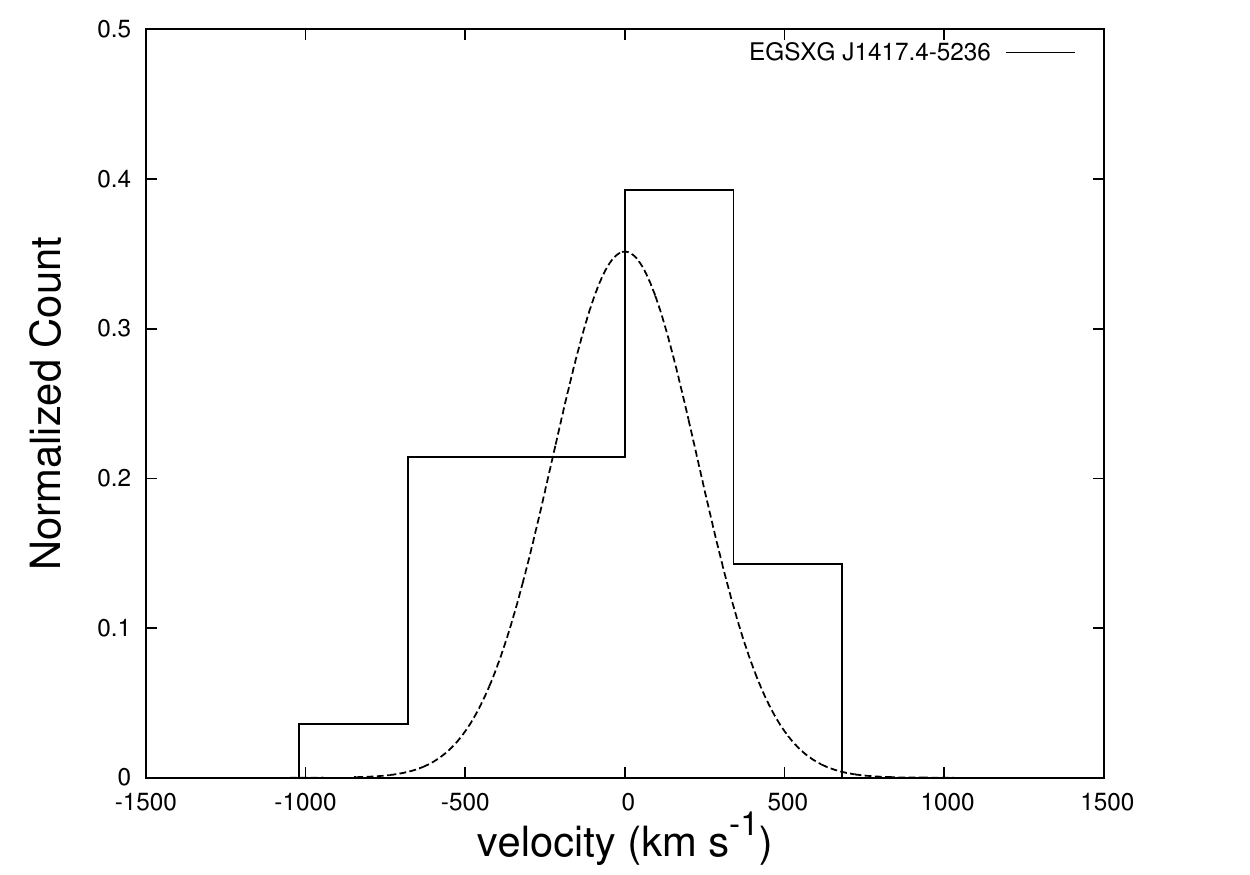}

\figcaption{The solid line shows the histogram of velocity distribution of EGSXG J1417.3+5235. The dashed Gaussian curve is the expected velocity distribution from scaling relation for this group.\label{fig111}}

\end{figure}

 In the cases of EGSXG J1417.3+5235 and EGSXG J1417.7+5241, both having large numbers of spectroscopic member galaxies, the estimation of velocity dispersion can't be the main uncertainty. 
 EGSXG J1417.3+5235 satisfies all the three criteria (population, isolation and compactness) for a compact group (\citealt{Hic82}). For compactness criterion, \citet{Hic82}
establish that the sum of member galaxies' magnitude averaged over the smallest circle containing the cores of most luminous galaxies in a compact group should be less than 26 in POSS-I E band,
 $\mu_E< 26$ mag arcsec$^{-2}$. He use POSS-I E band for the cut on the surface brightness of his local groups which roughly corresponds to r-band (e.g. \citealt{Dia12}).
 As all his compact groups are in the local universe, we should apply the $k$-correction to the r-band magnitude of the member galaxies of EGSXG J1417.3+5235 at z=0.236.
 All galaxies which we use for computing surface brightness are on the red-sequence, therefore, we calculate the $k$-correction to the r-band magnitude using the stellar population model
 of \citet{Mar09} for red galaxies. Using the $k$-corrected magnitudes, this group satisfies compactness criterion. It also has high 
concentration in X-ray emission (Figure \ref{fig13}), while having low mass, so leading to steep X-ray profile. However, \citet{Hel00} find the loose and compact local groups lie in a
 similar position on the $L_X-\sigma$ relation. Figure \ref{fig111} shows the histogram of velocity distribution of member galaxies and the expected Gaussian distribution from scaling
 relations for EGSXG J1417.3+5235.

 Excluding these three groups, the Flag=2 groups and also the group with substructures, the $L_x-\sigma$ relation of our sample is consistent with the $L_x-\sigma$ relation expected from scaling
 relations obtained from COSMOS (\citealt{Lea10}). 


\begin{figure*}[t]

\centering
 \includegraphics[width=15cm]{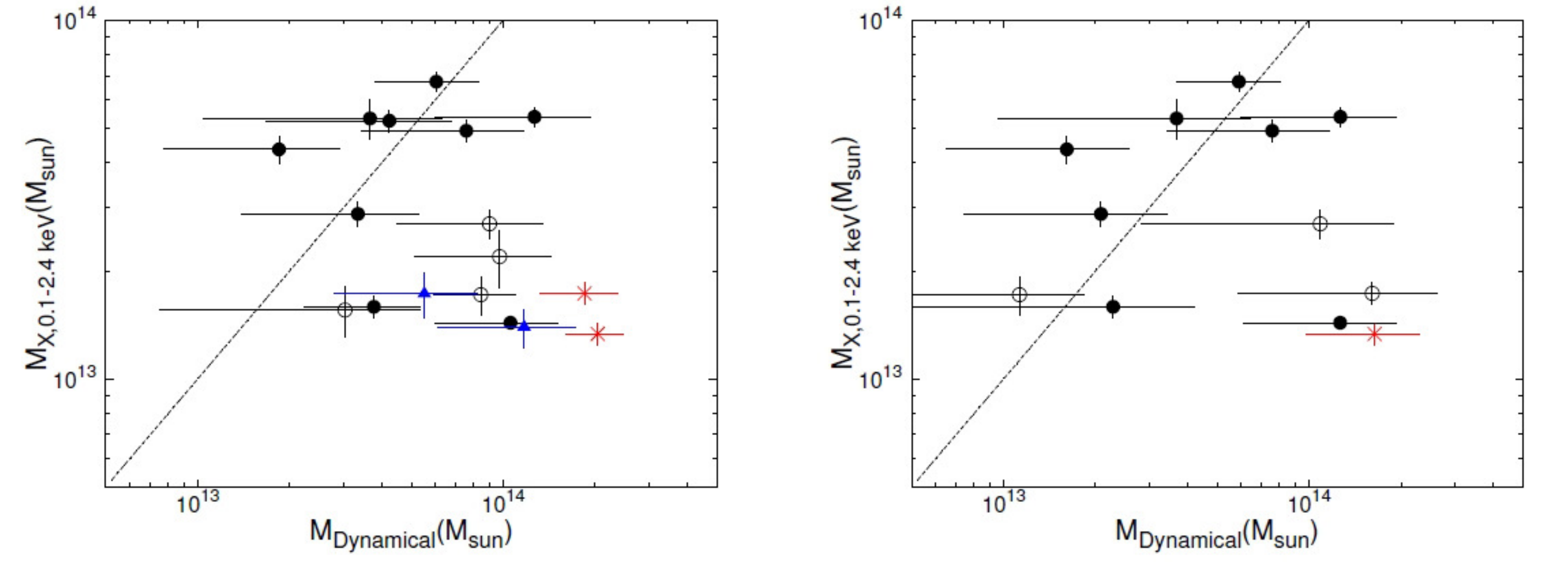}

 \figcaption{X-ray mass versus dynamical mass for the X-ray galaxy groups. Left plot shows the galaxy groups with member selected within velocity dispersion based $r_{200}$. The right one shows galaxy groups with 
members within X-ray based $r_{200}$. The red stars show groups which have substructure detected by DS test, blue triangles are the compact X-ray systems and open circles show the 
groups with Flag=2. The dashed line is the one-to-one relation.\label{fig14}} 
 \end{figure*}

\subsection{X-ray mass vs. Dynamical mass}
We also estimated dynamical mass for the galaxy groups using $r_{200}$  (eq. 10) and the intrinsic velocity dispersion as in \citet{Bal06} and \citet{Carl99}:

 \begin{equation}
 \label{eqn:11}
 M_{dyn}=\frac{3}{G} \sigma^2 r_{200}
\end{equation}
As we expect from the $L_x-\sigma$ relation of the groups (Figures \ref{fig10} and \ref{fig11}) we find much better agreement between dynamical mass and X-ray mass for high mass systems (Figure \ref{fig14}).
 For low mass systems with low quality flag (Flag=2) and substructure, the disagreement between these two masses is more substantial. 
In addition, the errors on dynamical mass are increased in X-ray based $r_{200}$ as the groups have less members in this case compared to that of an optically based $r_{200}$.

\section{Comparison to optical groups}

\begin{figure}[t]
 \includegraphics[width=9cm]{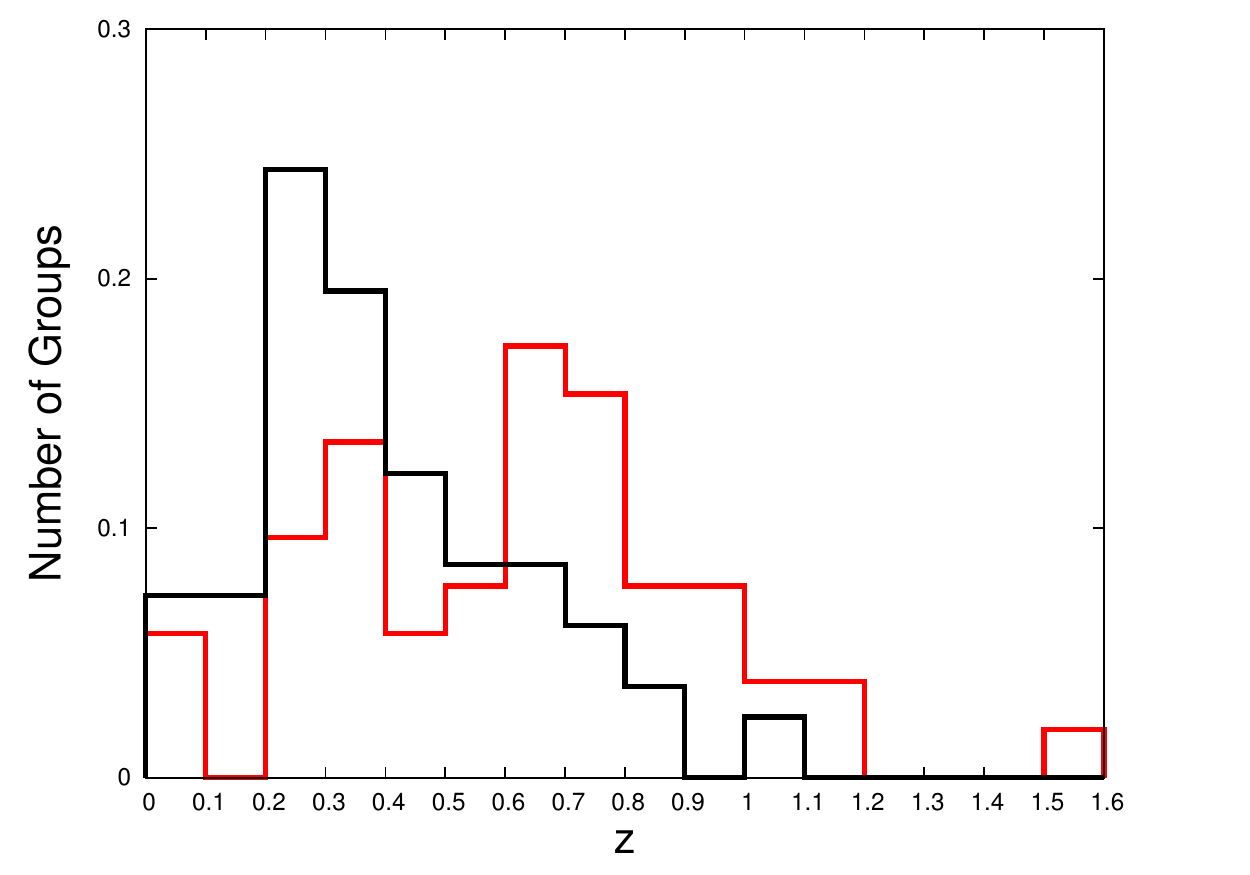}
\figcaption{Normalized distribution of redshift for X-ray galaxy groups (in red) and optical galaxy groups (in black) inside of our flux detection limits in EGS.\label{fig15}}
\end{figure}

The optical group catalog is derived from DEEP2 DR4 dataset using the Voronoi-Delaunay 
method (VDM) group finder (\citealt{Ger12}) and includes groups in all DEEP2 fields. It yields 1165 groups 
with more than two observed members in the EGS field. 
We look at the distribution of redshifts for our X-ray and optically selected groups and in Figure \ref{fig15} show the normalized distribution of redshifts for both samples. X-ray groups are preferentially found at z$>$0.6, compared to 
optical groups, but in general the two distributions are similar. 

\begin{figure}[t]
 \includegraphics[width=9cm]{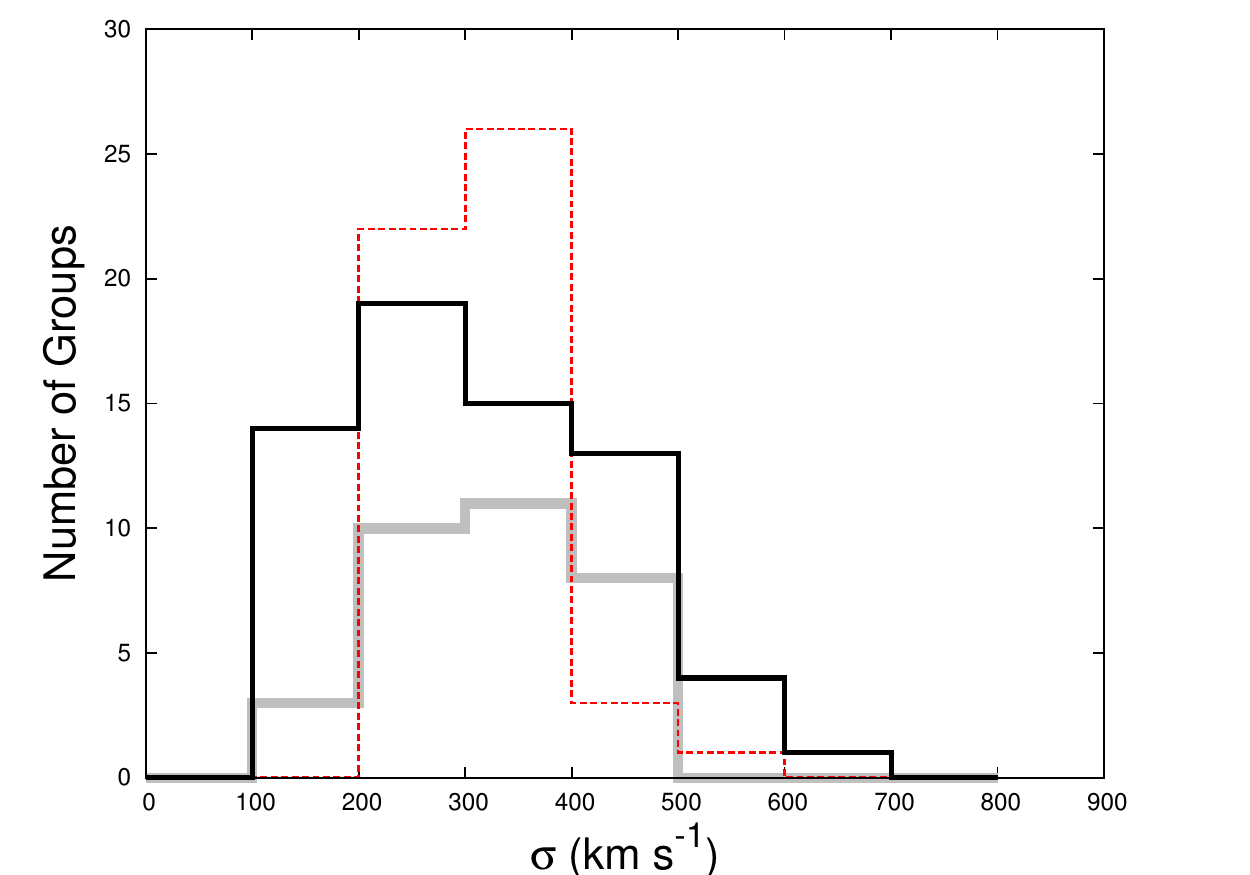}
 \figcaption{Distribution of velocity dispersion for optical and X-ray galaxy groups. The black line shows the distributions of velocity dispersion of optical VDM groups. 
The red and thick grey lines are corresponding
to distributions of velocity dispersion estimated from X-rays and dynamical velocity dispersion for X-rays groups.\label{fig16} }
 \end{figure}

We also compared velocity dispersions of both samples. In order to have a reliable comparison we take into account only optical groups with our X-ray detection limit. Figure \ref{fig16} 
shows the distribution of rest-frame velocity dispersion for galaxy groups with more than five members. For X-ray galaxy groups, we plot both velocity dispersions derived from X-ray properties and gapper 
estimator method. Velocity dispersion of optical groups are the observed velocity dispersion derived using the gapper algorithm. We converted them to rest-frame velocity using Eq. 7 for the comparison. 

We searched for the high velocity tail of the distribution of the optical group velocity dispersion in the X-ray data. This high velocity tail arises from galaxy groups with
 less than ten members which are actually part of a bigger group or on the edge of the X-ray coverage of EGS. Ignoring this tail, in general these distributions are also similar.\\

\section{Summary and Discussion}
We have searched for X-ray extended sources in AEGIS. We identified X-ray galaxy groups corresponding to extended X-ray emissions and presented the galaxy groups catalog. 
 The catalog is reaching fluxes of $10^{-15}$ ergs s$^{-1}$ cm$^{-2}$ with 52 systems detected by Chandra. 
Previously Chandra catalogs at such depths contain half a dozen objects per field (\citealt{Gia02, Bau02}). The DEEP2 and DEEP3 redshift surveys which provide the most complete sample at intermediate 
redshift and the largest accurate data sets at $z\sim1$, combined with deep X-ray imaging from Chandra in EGS field, bring a deep study of galaxy groups to a new level.  
 Spectroscopic member galaxies are selected by applying different cuts:
 two X-ray based and optically based virial radius and two cuts along the line of sight. We examined the $L_x-\sigma$ relation for each cut and discussed the effects
 of them on the relation. We explored substructure in the groups by applying the Dressler-Shectman test and discussed its effect on the overestimation of velocity dispersion.
 We also looked at the compactness of the X-ray emission of the groups and its effect on the scaling relation. A comparison between dynamical mass and X-ray mass of the groups is also done.
 Finally, X-ray galaxy groups are compared with optical galaxy groups which are identified from VDM groups.
Our results show:\\
1) Our detection of a high-z group illustrates that megasecond  Chandra exposures are required for detecting such objects in the volume of deep fields. Smaller exposures, as
 \citet{Pie12}, only yields a marginal detection.
2) For a sample of groups with a wide range of X-ray luminosities, choise of a constant $\Delta z$ for member selection can cause a large scatter in $L_x-\sigma$ relation. 
3) Choice of X-ray based virial radius or optically based virial radius does not have significant effect on the scatter around $L_x-\sigma$ relation for groups with high X-ray luminosities.  
4) Substructure in groups can inflate velocity dispersion as outliers are also included in galaxies with dynamical complexity. 
5) We find a compact group at z=0.24 with a concentrated X-ray emission with high velocity dispersion in comparison to X-ray luminosity.
6) X-ray galaxy groups and optical galaxy groups from VDM are nearly similar in both redshift and velocity dispersion distributions.
\\

 We would like to thank David Wilman, Annie Hou and Antonis Georgakakis for helpful discussions about the dynamical properties and AGN contamination of X-ray groups.
 We also thank Felicia Ziparo and Steve Willner for useful comments on the manuscript.
MT acknowledges the support by World Premier International Research Center Initiative (WPI Initiative), MEXT, Japan and by KAKENHI No. 23740144.
 This work was also supported in part by National Science Foundation grants AST 00-71198, 05-07483, 08-08133 to UCSC and AST 00-71048, 05-07428, and 08-07630 to UCB and AST 08-06732 to U. Pittsburgh.
 This project has been supported by the DLR grant 50OR1013 to MPE.

\clearpage

\clearpage

\clearpage

\clearpage

\clearpage

\clearpage

\clearpage

\clearpage

\clearpage

\clearpage

\clearpage
\begin{deluxetable}{lccccccccccc}
\tabletypesize{\scriptsize}
\rotate
\tablecaption{AEGIS X-ray Galaxy Groups}
\tablewidth{0pt}
\tablehead{
\colhead{ID} & \colhead{RA} & \colhead{DEC} & \colhead{z} & \colhead{Flux} &
\colhead{L$_X$} & \colhead{M$_{200}$} & \colhead{r$_{200}$} &\colhead{Flag} & \colhead{N(z)} &
\colhead{Flux} & \colhead{$\sigma_{X}$}\\
\colhead{}&\colhead{}&\colhead{}&\colhead{}&\colhead{[$10^{-14}ergcm^{-2}s^{-1}$]}&\colhead{[$10^{42}erg s^{-1}$]}&\colhead{[$10^{13}M_\odot$]}&\colhead{[arcmin]}&\colhead{}&\colhead{}&\colhead{Significance}&\colhead{}}


\startdata
 EGSXG J1414.8+5209 & 213.71657 & 52.16481 & 0.455 & 0.50$\pm$0.08 & 6.35$\pm$0.98 & 4.61$\pm$0.44 & 1.8 & 4 & 3 & 6.50 & 325 \\
 EGSXG J1414.8+5212 & 213.71946 & 52.21343 & 0.301 & 0.38$\pm$0.09 & 1.76$\pm$0.43 & 2.32$\pm$0.35 & 2.0 & 2 & 3 & 4.11 & 251\\
 EGSXG J1415.0+520$4^\dag$ & 213.76555 & 52.07685 & 0.074 & 2.77$\pm$0.33 & 0.56$\pm$0.07 & 1.34$\pm$0.10 & 5.7 & 1 & 28 & 8.49 & 202 \\
 EGSXG J1415.4+522$0^\dag$ & 213.85106 & 52.34112 & 0.074 & 4.22$\pm$0.46 & 0.84$\pm$0.09 & 1.73$\pm$0.12 & 6.2 & 2 & 19 & 9.19 & 220\\
 EGSXG J1415.6+5221 & 213.91018 & 52.36540 & 0.622 & 0.33$\pm$0.07 & 9.20$\pm$1.94 & 5.01$\pm$0.65 & 1.5 & 4 & 9 & 4.73 & 345 \\
 EGSXG J1416.1+5224 & 214.02791 & 52.41598 & 0.571 & 0.13$\pm$0.07 & 3.19$\pm$1.67 & 2.67$\pm$0.82 & 1.3 & 1 & 6 & 1.92 & 277\\
 EGSXG J1416.2+520$5^\dag$ & 214.07123 & 52.09987 & 0.832 & 0.91$\pm$0.13 & 46.87$\pm$6.79 & 11.68$\pm$1.06 & 1.6 & 2 & 1 & 6.90 & 475\\
 EGSXG J1416.3+5213 & 214.07507 & 52.22752 & 0.641 & 0.12$\pm$0.04 & 4.02$\pm$1.46 & 2.90$\pm$0.64 & 1.2 & 4 & 7 & 2.76 & 288\\
 EGSXG J1416.3+5229 & 214.07862 & 52.49943 & 0.356 & 0.36$\pm$0.08 & 2.46$\pm$0.55 & 2.74$\pm$0.38 & 1.8 & 2 & 8 & 4.44 & 268\\
 EGSXG J1416.3+5214 & 214.08759 & 52.23544 & 0.366 & 0.25$\pm$0.06 & 1.83$\pm$0.43 & 2.25$\pm$0.32 & 1.7 & 4 & 8 & 4.27 & 251 \\
 EGSXG J1416.4+5227 & 214.12227 & 52.45173 & 0.837 & 0.30$\pm$0.06 & 17.81$\pm$3.26 & 6.26$\pm$0.71 & 1.3 & 1 & 7 & 5.46 & 386\\
 EGSXG J1416.6+5228 & 214.15991 & 52.47882 & 0.812 & 0.09$\pm$0.04 & 5.94$\pm$2.34 & 3.17$\pm$0.75 & 1.1 & 3 & 5 & 2.54 & 307\\
 EGSXG J1416.6+5222 & 214.17417 & 52.37189 & 0.510 & 0.15$\pm$0.04 & 2.63$\pm$0.79 & 2.50$\pm$0.45 & 1.4 & 2 & 5 & 3.35 & 267 \\
 EGSXG J1416.6+5229 & 214.17480 & 52.48355 & 0.238 & 0.33$\pm$0.09 & 0.87$\pm$0.23 & 1.56$\pm$0.25 & 2.1 & 4 & 6 & 3.82 & 218\\
 EGSXG J1416.8+5210 & 214.20403 & 52.17700 & 0.900 & 1.04$\pm$0.15 & 63.55$\pm$8.95 & 13.33$\pm$1.17 & 1.6 & 3 & 0 & 7.10 & 503 \\
 EGSXG J1417.0+5226 & 214.25416 & 52.44758 & 1.023 & 0.08$\pm$0.02 & 10.89$\pm$2.90 & 3.85$\pm$0.63 & 1.0 & 1 & 5 & 3.75 & 340 \\
 EGSXG J1417.2+5215 & 214.31665 & 52.25140 & 0.470 & 0.20$\pm$0.05 & 2.82$\pm$0.72 & 2.71$\pm$0.42 & 1.5 & 3 & 0 & 3.95 & 273\\
 EGSXG J1417.3+5235 & 214.34115 & 52.59349 & 0.236 & 0.27$\pm$0.05 & 0.73$\pm$0.14 & 1.39$\pm$0.17 & 2.1 & 1 & 9 & 5.07 & 210\\
 EGSXG J1417.4+5237 & 214.37150 & 52.63047 & 0.355 & 0.39$\pm$0.05 & 2.66$\pm$0.34 & 2.88$\pm$0.23 & 1.9 & 1 & 14 & 7.72 & 273\\
 EGSXG J1417.5+5238 & 214.38305 & 52.63655 & 0.717 & 0.29$\pm$0.06 & 11.61$\pm$2.36 & 5.32$\pm$0.67 & 1.4 & 2 & 13 & 4.93 & 358 \\
 EGSXG J1417.5+5232 & 214.38819 & 52.53527 & 0.985 & 0.25$\pm$0.03 & 22.59$\pm$2.87 & 6.35$\pm$0.51 & 1.2 & 2 & 6 & 7.87 & 399 \\
 EGSXG J1417.7+5228 & 214.44405 & 52.47140 & 0.995 & 0.16$\pm$0.04 & 16.36$\pm$4.08 & 5.12$\pm$0.79 & 1.1 & 1 & 5 & 4.01 & 372 \\
 EGSXG J1417.7+524$1^\dag$ & 214.44751 & 52.69237 & 0.066 & 3.92$\pm$0.24 & 0.62$\pm$0.04 & 1.43$\pm$0.06 & 6.5 & 1 & 23 & 16.27 & 206 \\
 EGSXG J1417.9+5226 & 214.47592 & 52.43701 & 0.684 & 0.13$\pm$0.03 & 5.04$\pm$1.30 & 3.22$\pm$0.51 & 1.2 & 4 & 3 & 3.88 & 301 \\
 EGSXG J1417.9+5235 & 214.48176 & 52.58382 & 0.683 & 0.27$\pm$0.03 & 9.76$\pm$1.04 & 4.92$\pm$0.33 & 1.4 & 1 & 11 & 9.42 & 346\\
 EGSXG J1417.9+5225 & 214.48632 & 52.42462 & 0.995 & 0.10$\pm$0.03 & 11.68$\pm$3.58 & 4.13$\pm$0.77 & 1.0 & 3 & 5 & 3.26 & 347\\
 EGSXG J1417.9+5231 & 214.49046 & 52.52231 & 0.661 & 0.16$\pm$0.03 & 5.52$\pm$0.93 & 3.49$\pm$0.37 & 1.3 & 1 & 6 & 5.91 & 308\\
 EGSXG J1418.0+5222 & 214.50330 & 52.36890 & 0.950 & 0.22$\pm$0.04 & 18.79$\pm$3.81 & 5.83$\pm$0.73 & 1.2 & 3 & 0 & 4.94 & 386\\
 EGSXG J1418.1+5225 & 214.52697 & 52.41944 & 0.548 & 0.21$\pm$0.05 & 4.40$\pm$0.96 & 3.35$\pm$0.45 & 1.4 & 1 & 3 & 4.59 & 297 \\
 EGSXG J1418.3+5227 & 214.59199 & 52.45022 & 0.281 & 0.27$\pm$0.06 & 1.08$\pm$0.25 & 1.73$\pm$0.24 & 1.9 & 2 & 7 & 4.36 & 227 \\
 EGSXG J1418.5+5252 & 214.62561 & 52.86899 & 1.025 & 0.10$\pm$0.03 & 12.31$\pm$3.83 & 4.15$\pm$0.79 & 1.0 & 2 & 1 & 3.21 & 349 \\
 EGSXG J1418.8+5248 & 214.70147 & 52.80696 & 0.741 & 0.06$\pm$0.02 & 3.37$\pm$1.07 & 2.36$\pm$0.45 & 1.0 & 1 & 4 & 3.15 & 274 \\
 EGSXG J1419.0+5236 & 214.76363 & 52.61357 & 0.550 & 0.45$\pm$0.04 & 9.25$\pm$0.85 & 5.38$\pm$0.31 & 1.7 & 1 & 11 & 10.86 & 348 \\
 EGSXG J1419.0+5257 & 214.76810 & 52.95626 & 0.656 & 0.32$\pm$0.04 & 10.37$\pm$1.15 & 5.24$\pm$0.36 & 1.5 & 1 & 16 & 9.02 & 352 \\
 EGSXG J1419.2+5255 & 214.82121 & 52.91803 & 0.782 & 0.14$\pm$0.02 & 7.51$\pm$1.29 & 3.79$\pm$0.41 & 1.2 & 3 & 10 & 5.80 & 324 \\
 EGSXG J1419.4+5301 & 214.85110 & 53.02358 & 0.745 & 0.15$\pm$0.03 & 6.95$\pm$1.52 & 3.74$\pm$0.50 & 1.2 & 1 & 6 & 4.57 & 320\\
 EGSXG J1419.6+5251 & 214.91782 & 52.85103 & 0.670 & 0.14$\pm$0.03 & 4.99$\pm$1.24 & 3.24$\pm$0.50 & 1.2 & 1 & 6 & 4.01 & 301\\
 EGSXG J1419.7+5246 & 214.93393 & 52.77711 & 0.784 & 0.18$\pm$0.06 & 9.53$\pm$3.10 & 4.41$\pm$0.87 & 1.2 & 1 & 6 & 3.07 & 340 \\
 EGSXG J1419.8+5300 & 214.96203 & 53.01138 & 0.744 & 0.19$\pm$0.03 & 8.85$\pm$1.29 & 4.36$\pm$0.40 & 1.3 & 1 & 14 & 6.87 & 337\\
 EGSXG J1420.0+5258 & 215.00399 & 52.97709 & 0.454 & 0.15$\pm$0.03 & 1.91$\pm$0.39 & 2.14$\pm$0.27 & 1.4 & 1 & 7 & 4.95 & 251 \\
 EGSXG J1420.0+5306 & 215.00462 & 53.11241 & 0.200 & 0.48$\pm$0.05 & 0.85$\pm$0.10 & 1.59$\pm$0.11 & 2.5 & 2 & 19 & 8.93 & 218\\
 EGSXG J1420.2+5308 & 215.06885 & 53.13864 & 1.105 & 0.13$\pm$0.04 & 18.62$\pm$5.14 & 5.03$\pm$0.85 & 1.0 & 2 & 3 & 3.63 & 378\\
 EGSXG J1420.4+5311 & 215.12046 & 53.19035 & 1.544 & 0.18$\pm$0.04 & 54.10$\pm$13.11 & 6.80$\pm$1.01 & 0.9 & 2 & 3 & 4.13 & 451 \\
 EGSXG J1420.5+530$8^\dag$ & 215.14733 & 53.13952 & 0.734 & 0.41$\pm$0.04 & 17.19$\pm$1.75 & 6.74$\pm$0.43 & 1.5 & 1 & 17 & 9.84 & 388 \\
 EGSXG J1420.8+5306 & 215.20148 & 53.10867 & 0.355 & 0.35$\pm$0.05 & 2.41$\pm$0.36 & 2.71$\pm$0.25 & 1.8 & 2 & 14 & 6.73 & 267 \\
 EGSXG J1421.3+5308 & 215.32631 & 53.14764 & 0.351 & 0.26$\pm$0.08 & 1.72$\pm$0.51 & 2.19$\pm$0.39 & 1.7 & 2 & 7 & 3.40 & 249 \\
 EGSXG J1421.4+5308 & 215.34650 & 53.13477 & 0.201 & 0.54$\pm$0.11 & 0.97$\pm$0.20 & 1.72$\pm$0.21 & 2.5 & 2 & 15 & 4.96 & 224\\
 EGSXG J1421.5+5302 & 215.38609 & 53.04315 & 0.677 & 0.53$\pm$0.18 & 17.62$\pm$6.11 & 7.22$\pm$1.52 & 1.6 & 2 & 5 & 2.88 & 393  \\
 EGSXG J1422.0+5328  & 215.51144 & 53.47268 & 0.357 & 0.51$\pm$0.19 & 3.54$\pm$1.30 & 3.46$\pm$0.77 & 2.0 & 1 & 6 & 2.71 & 290\\
 EGSXG J1422.6+532$1^\dag$ & 215.66020 & 53.36037 & 0.768 & 0.36$\pm$0.10 & 16.95$\pm$4.73 & 6.47$\pm$1.10 & 1.4 & 1 & 9 & 3.58 & 388  \\
 EGSXG J1423.0+5326 & 215.76783 & 53.44306 & 0.615 & 0.28$\pm$0.11 & 7.66$\pm$2.94 & 4.49$\pm$1.04 & 1.5 & 2 & 3 & 2.60 & 332 \\
 EGSXG J1423.6+532$8^\dag$ & 215.92348 & 53.47078 & 1.132 & 0.27$\pm$0.05 & 34.88$\pm$6.68 & 7.34$\pm$0.87 & 1.2 & 1 & 5 & 5.22 & 431\\
\enddata

$^\dag$ Galaxy groups previously reported in Jeltema et al. (2009)
\end{deluxetable}
\clearpage

\begin{table}
\begin{center}
\caption{A sample of group member galaxies.\label{tbl-2}}
\begin{tabular}{lccc}
\hline
\hline
Group ID & RA & Dec. & z  \\
\hline
EGSXG J1417.5+5238& 214.38009& 52.63223&	0.7199 \\  
EGSXG J1417.5+5238& 214.36919& 52.62006&	0.7201 \\  
EGSXG J1417.5+5238& 214.36626& 52.65201&	0.7157 \\   
EGSXG J1417.5+5238& 214.41671& 52.62889&	0.7173 \\ 
EGSXG J1417.5+5238& 214.40832& 52.62752&	0.7193 \\  
EGSXG J1417.5+5238& 214.40229& 52.62471&	0.7170 \\  
EGSXG J1417.5+5238& 214.40131& 52.63039&	0.7168 \\ 
EGSXG J1417.5+5238& 214.39906& 52.64248&	0.7151 \\ 
EGSXG J1417.5+5238& 214.39653& 52.64163&	0.7161 \\  
EGSXG J1417.5+5238& 214.39236& 52.62690&	0.7167 \\     
EGSXG J1417.5+5238& 214.39221& 52.63439&	0.7156 \\   
EGSXG J1417.5+5238& 214.38562& 52.63828&	0.7168 \\   
EGSXG J1417.5+5238& 214.34541& 52.63622&	0.7167 \\   
EGSXG J1419.6+5251& 214.90556& 52.85547&	0.6675 \\ 
EGSXG J1419.6+5251& 214.94009& 52.85377&	0.6713 \\ 
EGSXG J1419.6+5251& 214.92459& 52.85621&	0.6711 \\ 
EGSXG J1419.6+5251& 214.91483& 52.84952&	0.6692 \\ 
EGSXG J1419.6+5251& 214.90534& 52.85083&	0.6702 \\ 
EGSXG J1419.6+5251& 214.90136& 52.85234&	0.6681 \\ 
		
\hline
\tablecomments{Table is published in its entirety in the electronic edition of the Astrophysical Journal. A portion is shown here for guidance regarding its form and content.}
\end{tabular}
\end{center}
\end{table}


\begin{thebibliography}{}
\bibitem[Balogh et al.(2006)]{Bal06} Balogh, M. L., Babul, A., Voit, G. M, McCarthy, I. G., Jones, L. R., Lewis, G. F.,\& Ebeling, H. 2006, \mnras, 366, 624 
\bibitem[Bauer et al.(2002)]{Bau02} Bauer, F. E., Alexander, D. M., Brandt, W. N., et al., 2002, \aj, 123, 1163B
\bibitem[Beers et al.(1990)]{Bee90} Beers, T. C., Flynn, K., Gebhardt, K., 1990, \aj, 100, 32B 
\bibitem[Benson et al.(2004)]{Ben04}  Benson, B. A., Church, S. E., Ade, P. A. R., Bock, J. J., Ganga, K. M., Henson, C. N., Thompson, K. L., 2004, \apj, 617, 829B 
\bibitem[Bielby et al.(2010)]{Bie10} Bielby, R. M., Finoguenov, A., Tanaka, M., McCracken, H. J., et al., 2010, \aap, 523A, 66B 
\bibitem[Bielby et al.(2012)]{Bie12} Bielby, R.; Hudelot, P.; McCracken, H. J., et al. 2012, \aap, 545A, 23B;
\bibitem[Boehringer et al.(2000)]{Boe00}  Boehringer, H., Voges, W., Huchra, J. P., McLean, B., et al., 2000, \apjs, 129, 435B 	
\bibitem[Brimioulle et al.(2008)]{Bri08}  Brimioulle, F. , Lerchster, M. , Seitz, S. , Bender, R. , Snigula, J. , 2008, arXiv, 0811.3211B 
\bibitem[Carlberg et al.(1999)]{Carl99}  Carlberg, R. G., et al. 1999, Royal Society of London Philosophical Transactions Series A, 357, 167 
\bibitem[Carlberg et al.(1997)]{Carl97} Carlberg, R. G., Yee, H. K. C., \& Ellingson, E. 1997, \apj, 478, 462
\bibitem[Carlstrom et al.(2002)]{Carl02} Carlstrom, J. E., Holder, G. P., Reese, E. D., 2002, \araa, 40, 643C
\bibitem[Coil et al.(2009)]{Coil09}  Coil, A. L., Georgakakis, A., Newman, J. A., et al. , 2009, \apj, 701, 1484C 
\bibitem[Connelly et al.(2012)]{Con12}  Connelly, J. L., Wilman, David J., Finoguenov, Alexis, et al. 2012, \apj, 756, 139C
\bibitem[Cooper et al.(2011)]{Coo11} Cooper, M. C., Aird, J. A., Coil, A. L., et al., 2011, \apjs, 193, 14C 
\bibitem[Cooper et al.(2012)]{Coo12} Cooper, M. C., Griffith, R. L. , Newman, J. A., Coil, A. L., et al., 2012, \mnras, 419, 3018C					   
\bibitem[Croston et al.(2005)]{Cro05}  Croston J. H., Hardcastle M. J., Harris D. E., Belsole E., Birkinshaw M., Worrall D. M., 2005, \apj, 626, 733 
\bibitem[Davis et al.(2003)]{Dav03}  Davis, M. , Faber, S. M., Newman, J. , et al., 2003, SPIE, 4834, 161D 
\bibitem[Davis et al.(2007)]{Dav07}  Davis, M., Guhathakurta, P., Konidaris, N. P., et al., 2007, \apj, 660L, 1D 
\bibitem[D\'{\i}az-Gim\'{e}nez et al.(2012)]{Dia12} D\'{\i}az-Gim\'{e}nez, E., Mamon, G. A., Pacheco, M., Mendes de Oliveira, C., Alonso, M. V., 2012, \mnras, 426, 296D
\bibitem[Dressler \& Shectman(1988)]{Dre88}  Dressler, A., \& Shectman, S. A. 1988, \aj, 95, 985
\bibitem[Efron (1982)]{Efr82} Efron, B. 1982, The Jackknife, the Bootstrap, and Other Resampling Plans (CBMS-NSF Regional Conference Series in Applied Mathematics; Philadelphia:
SIAM)
\bibitem[Eisenhardt et al.(2008)]{Eis08}  Eisenhardt, P. R. M., Brodwin, M., Gonzalez, A. H., et al., 2008, \apj, 684, 905E 
\bibitem[Eke et al.(2005)]{Eke05} Eke, V. R., Baugh, C. M., Cole, S., Frenk, C. S., King, H. M., \& Peacock, J. A. 2005, \mnras, 362, 1233
\bibitem[Faber et al.(2003)]{Fab03} Faber, S. M., Phillips, A. C., Kibrick, R., I., et al., 2003, SPIE, 4841, 1657F 
\bibitem[Finoguenov et al.(2007)]{Fin07} Finoguenov, A., Guzzo, L. , Hasinger, G., et al. 2007, \apjs, 172, 182F 
\bibitem[Finoguenov et al.(2009)]{Fin09} Finoguenov, A., Connelly, J. L., Parker, L. C., et al. 2009, \apj, 704, 564F 
\bibitem[Finoguenov et al.(2010)]{Fin10} Finoguenov, A., Watson, M. G., Tanaka, M., et al. 2010, \mnras, 403, 2063F  
\bibitem[Geller et al.(1983)]{Gel83} Geller, M. J.,\& Huchra, J. P. 1983, \apjs, 52, 61 
\bibitem[Gerke et al.(2012)]{Ger12} Gerke, B. F.; Newman, J. A.; Davis, M., et al., 2012, \apj, 751, 50G 
\bibitem[Giacconi et al.(2002)]{Gia02} Giacconi, R., Zirm, A., Wang, J., et al., 2002, \apjs, 139, 369G
\bibitem[Gilbank et al.(2011)]{Gil11} Gilbank, David G.; Gladders, M. D.; Yee, H. K. C.; Hsieh, B. C., 2011, \aj, 141, 94G
\bibitem[Girardi \& Mezzetti(2001)]{Gir01}  Girardi, M., \& Mezzetti, M. 2001, \apj, 548, 79
\bibitem[Gladders \& Yee (2005)]{Gla05} Gladders, M. D.,\& Yee, H. K. C. ,2005, \apjs, 157, 1G
\bibitem[Hasinger et al.(2001)]{Has01} Hasinger, G., Altieri, B., Arnaud, M.,et al., 2001, \aap, 365L, 45H 
\bibitem[Helsdon \& Ponman (2000)]{Hel00} Helsdon, S. F.; Ponman, T. J., 2000, \mnras, 319, 933H
\bibitem[Hickox \& Markevitch (2006)]{Hic06} Hickox, R. C.,\& Markevitch, M., 2006, \apj, 645, 95H
\bibitem[Hickson (1982)]{Hic82} Hickson, P., 1982, \apj, 259, 930H
\bibitem[Hou et al.(2012)]{Hou12} Hou, Annie, Parker, Laura C., Wilman, David J., et al., 2012, \mnras, 421, 3594H
\bibitem[Jeltema et al.(2009)]{Jel09}  Jeltema, T. E., Gerke, B. F., Laird, E. S., et al. 2009, \mnras, 399, 715J 
\bibitem[Jones et al. (2000b)]{Jon00} Jones, L. R., Ponman, T. J., Forbes, Duncan A., 2000, \mnras, 312, 139
\bibitem[Jones et al. (2003)]{Jon03}  Jones, L. R., Ponman, T. J., Horton, A., Babul, A., Ebeling, H.. Burke, D. J., 2003, \mnras, 343, 627J
\bibitem[Khosroshahi et al. (2007)]{sha07} Khosroshahi, H. G., Ponman, T. J., Jones, L. R., 2007, \mnras, 377, 595K
\bibitem[Knobel et al.(2009)]{Kno09}  Knobel, C., Lilly, S. J., Iovino, A., Porciani, C.,et al., 2009, \apj, 697, 1842K 
\bibitem[Koester et al.(2007)]{Koe07}  Koester, B. P., McKay, T. A., Annis, J., Wechsler, R. H., et al. 2007, \apj, 660, 221K
\bibitem[Laird et al.(2009)]{Lai09} Laird, E. S., Nandra, K., Georgakakis, A., Aird, J. A., et al. 2009, \apjs, 180,102L
\bibitem[LaRoque et al.(2003)]{Lar03}  LaRoque, S. J., Joy, M., Carlstrom, J. E., Ebeling H., Bonamente M., et al. 2003, \apj, 583, 559L
\bibitem[Leauthaud et al.(2010)]{Lea10}  Leauthaud, A., et al. 2010, \apj, 709, 97 
\bibitem[Maraston et al.(2009)]{Mar09}  Maraston, C; Str\"{o}mb\"{a}ck, G.; Thomas, D.; Wake, D. A.; Nichol, R. C., 2009, \mnras, 394L, 107M
\bibitem[Massey et al.(2007)]{Mas07}  Massey, R., Heymans, C., Bergé, J. ,et al., 2007, \mnras, 376, 13M 
\bibitem[Miller et al.(2005)]{Mil05}  Miller, C. J., Nichol, R. C., Reichart, D., et al., 2005, \aj, 130, 968M 
\bibitem[Miyazaki et al.(2007)]{Miy07}  Miyazaki, S., Hamana, T., Ellis, R. S., et al., 2007, \apj, 669, 714M 
\bibitem[Mulchaey \& Zabludoff (1998)]{Mul98}  Mulchaey, J. S.,\& Zabludoff, A. I., 1998, \apj, 496, 73M 
\bibitem[Newman et al.(2012)]{New12}  Newman, J. A., Cooper, M., Davis, M., Faber, S, et al., 2012, ArXiv:1203:3192
\bibitem[Osmond \& Ponman (2004)]{Osm04}  Osmond, J. P. F.; Ponman, T. J., 2004, \mnras, 350, 1511O
\bibitem[O'Sullivan \& Ponman (2004)]{Osu04} O'Sullivan, E.; Ponman, T. J., 2004, \mnras, 354., 935O
\bibitem[O'Sullivan et al. (2003)]{Osu03} O'Sullivan E. J., Ponman T. J., Collins R. S., 2003, \mnras, 340, 1375
\bibitem[Pierre  et al.(2012)]{Pie12}  Pierre, M., Clerc, N., Maughan, B., Pacaud, F., et al., 2012, \aap, 540A, 4P
\bibitem[Ponman et al.(1994)] {pon94}  Ponman, T. J., Allan, D. J., Jones, L. R., Merrifield, M., McHardy, I. M., Lehto, H. J., Luppino, G. A., 1994, \nat, 369, 462P
\bibitem[Stanford et al.(2006)]{Stan06}  Stanford, S. A., Romer, A. Kathy, Sabirli, Kivanc; et al., 2006, \apj, 646L, 13S 
\bibitem[Sunyaev \& Zeldovich(1970)]{Sun70}  Sunyaev, R. A.,\& Zeldovich, Ya. B., 1970, Ap\&SS, 7, 3S
\bibitem[Sunyaev \& Zeldovich(1972)]{Sun72}  Sunyaev, R. A.,\& Zeldovich, Ya. B., 1972, CoASP, 4, 173S
\bibitem[Staniszewski et al.(2009)]{Stan09}  Staniszewski Z., Ade P. A. R., Aird K. A.,et al. 2009, \apj, 701, 32S
\bibitem[Tanaka et al.(2010)]{Tan10}  Tanaka, M., Finoguenov, A., Ueda, Y., 2010, \apj, 716L, 152T
\bibitem[Turner \& Gott(1976)]{Tur76}  Turner, E. L., \& Gott, J. R.,III. 1976, \apjs, 32, 409
\bibitem[Vikhlinin et al.(2009)]{Vik09}  Vikhlinin, A., Burenin, R. A., Ebeling, H., et al., 2009, \apj, 692, 1033V 
\bibitem[Wilman et al.(2005)]{Wil05}  Wilman, D. J., Balogh, M. L., Bower, R. G.,et al., 2005, \mnras, 358, 71 
\bibitem[Zabludoff \& Mulchaey(1998)]{Zab98}  Zabludoff, A. I., \& Mulchaey, J. S., 1998, \apj, 496, 39

\end{thebibliography}
\end{document}